\documentclass[authoryear]{elsarticle}
\usepackage[english]{babel}

\usepackage[margin=0.6in]{geometry}

\usepackage[utf8]{inputenc}
\usepackage[bottom]{footmisc} 
\usepackage{graphicx}
\usepackage{rotating,booktabs}
\usepackage{natbib}
\usepackage{url}
\usepackage{booktabs}
\usepackage{graphicx}
\usepackage{adjustbox}

\usepackage{amsmath}
\usepackage{xcolor}
\usepackage{amsfonts}
\usepackage{amssymb}
\usepackage{graphicx}
\usepackage{subfigure}
\usepackage{pdflscape}
\usepackage{longtable}
\usepackage[small, justification=justified,singlelinecheck=false]{caption}
\usepackage{multirow}

\usepackage[figurename=Figure.,
justification=RaggedRight, 
labelfont={bf, footnotesize}, 
textfont={footnotesize},position=top]{caption}
\usepackage{color}
\definecolor{byzantium}{rgb}{0.44, 0.16, 0.39}	
\usepackage[english]{babel}
\usepackage{ulem}

\usepackage[margin=0.6in]{geometry}

\usepackage[utf8]{inputenc}
\usepackage[bottom]{footmisc} 
\usepackage{graphicx}
\usepackage{rotating,booktabs}
\usepackage{natbib}
\usepackage{url}
\usepackage{adjustbox}

\usepackage{amsmath}
\usepackage{amsfonts}
\usepackage{amssymb}
\usepackage{graphicx}
\usepackage{subfigure}
\usepackage{pdflscape}
\usepackage{longtable}
\usepackage[justification=justified,singlelinecheck=false]{caption}
\usepackage{multirow}
\usepackage[figurename=Figure.,
justification=RaggedRight, 
labelfont={bf, footnotesize}, 
textfont={footnotesize},position=top]{caption}
\usepackage{color}
\definecolor{byzantium}{rgb}{0.44, 0.16, 0.39}	
\usepackage{booktabs}
\usepackage{epstopdf}
\usepackage{setspace}
\usepackage[colorlinks=true]{hyperref}
\hypersetup{linkcolor=blue}
\hypersetup{citecolor=blue}
\usepackage[capposition=top]{floatrow}
\usepackage{setspace}
\usepackage{natbib}
\begin{document}

\begin{frontmatter}
\title{Anatomy of a Stablecoin’s failure: the Terra-Luna case}
		
\author[myaddress2,myaddress4]{Antonio Briola}
\author[myaddress2,myaddress,myaddress4]{David Vidal-Tom{\'a}s \corref{mycorrespondingauthor}}
\cortext[mycorrespondingauthor]{Corresponding author}
\ead{d.vidal@ucl.ac.uk}
\author[myaddress2,myaddress4]{Yuanrong Wang}
\author[myaddress2,myaddress3,myaddress4]{Tomaso Aste}

\address[myaddress2]{Department of Computer Science, University College London, Gower Street, WC1E 6EA - London, United Kingdom.}
\address[myaddress]{Department of Economics, Universitat Jaume I, Campus del Riu Sec, 12071 - Castellón, Spain.}
\address[myaddress3]{Systemic Risk Centre, London School of Economics, London, United Kingdom.}
\address[myaddress4]{UCL Centre for Blockchain Technologies, London, United Kingdom.}
	
\begin{abstract}

We quantitatively describe the main events that led to the Terra project’s failure in May 2022. We first review, in a systematic way, news from heterogeneous social media sources; we discuss the fragility of the Terra project and its vicious dependence on the Anchor protocol. We hence identify the crash’s trigger events, analysing hourly and transaction data for Bitcoin, Luna, and TerraUSD. Finally, using state-of-the-art techniques from network science, we study the evolution of dependency structures for 61 highly capitalised cryptocurrencies during the down-market and we also highlight the absence of herding behaviour analysing cross-sectional absolute deviation of returns.


\hfill

	
\end{abstract}
	

\end{frontmatter}

\pagebreak
	
\section{Introduction} \label{sec:Introduction}

In one of his public declarations, Terra project's co-founder, Do Kwon, stated that \textit{95\% of the companies} entering the crypto market were going to \textit{die}. Indeed, he further affirmed that there was \textit{entertainment in watching companies die} (\citealp{Quiroz-Gutierrez2022}). About a week after this statement, TerraUSD (UST), the stablecoin that, according to its founders, should have been the new peer-to-peer cash system, lost its peg to the U.S. dollar (USD) and collapsed.

Before its crash, UST was the fourth-largest stablecoin behind Tether (USDT), USD Coin (USDC) and Binance USD (BUSD), with $\$18$ billion in market capitalization (\citealp{UST_Cap}). According to Terra's white paper, its underlying protocol relied on a two-coin system, which was not fully backed by traditional collaterals (e.g. fiat currency or gold) (\citealp{kereiakes2019terra}). On the one hand, Terra was the algorithmic stablecoin whose value was pegged to different fiat currencies, giving rise to fiat-based stablecoins, such as TerraUSD, TerraEUR and TerraKRW. On the other hand, Luna token\footnote{Some coin-ranking sites, such as CoinMarketCap and Coingecko, use Terra and TerraUSD for the acronyms LUNA and UST. In this paper, to avoid any misconception, we denote Terra as the fiat-based stable coin and Luna as the counterpart.} (LUNA) was the counterweight used to delete (or, at least, reduce) the volatility from UST. Being more specific, LUNA-UST protocol was based on two main concepts. First, the protocol stabilised UST prices by ensuring that its supply and demand were in equilibrium through arbitrage, that is, contracting (or expanding) UST pool by using LUNA pool as counterweigth. Second, through the Terra protocol's algorithmic market module, arbitrageurs were allowed to trade \$1 worth of LUNA for 1 UST, and vice versa, regardless of LUNA and UST prices (\citealp{Shapovalov2022})\footnote{When UST price was traded below \$1 (e.g. \$0.99), it implied that supply for UST was higher than demand, thus, it was necessary to contract Terra pool. To rebalance supply and demand, arbitrageurs would have acquired 1 UST for \$0.99. Afterwards, they would have swapped 1 UST for \$1 of LUNA through the market module, obtaining the corresponding profit (e.g. \$0.01). Consequently, 1 UST would have been burned (decreasing its supply and contracting UST pool) and \$1 of LUNA would have been minted (increasing its supply and expanding LUNA pool). The arbitrage procedure would have continued, giving rise to an upward pressure on the UST price, due to its continuous supply decrease, until it would have reached the USD peg.}.

The most famous project born inside the Terra ecosystem was the Anchor protocol, a lending and borrowing protocol used by UST holders as an high-interest savings account. Before Terra project's collapse, Anchor attracted the 75\% of UST circulating supply offering 20\% annual percentage yield (APY) for depositors. Aided by its vicious dependence on the Anchor protocol, Terra project immediately appeared to be heavily exposed to extreme market conditions. On April 2022, Swiss-based crypto exchange, SwissBorg, stated that \textit{if LUNA’s price is under pressure, UST holders could be fearing that the UST peg is at risk and decide to redeem their UST positions. In order to do so, UST is burnt and LUNA is minted and sold on the market. This would exacerbate further the decline of LUNA’s price, pushing more UST holders to sell their UST. This vicious cycle is known as ``bank run" or ``death spiral"} (\citealp{SwissBorgReport}). This ``death spiral" effectively occurred to LUNA-UST on May 2022. Social media sources relates its origin to a ``coordinated attack", in which market actors strategically used their capital to destabilise the UST peg and generate profits (\citealp{Morris2022}). Specifically, two main events are identified as the fuse for the Terra collapse. First, private market actors short sold Bitcoin (BTC) with the final aim of spreading panic into the market (\citealp{Hall2022}; \citealp{Ashmore2022}; \citealp{Locke2022}). Second, on 07 May 2022, the liquidity pool Curve-3pool suffered a ``liquidity pool attack", which caused the first UST de-pegging, below \$0.99 (\citealp{Ashmore2022}; \citealp{Chainanalysis2022}). It is worth noting that, on 01 April 2022, \cite{Kwon2022b} announced the launch of a new liquidity pool (4pool) together with DeFi majors Frax Finance and Redacted Cartel. Consequently, funds were expected to move to the new pool, giving rise to a transitory situation in which Curve-3pool would have been more illiquid and prone to attacks. After the aforementioned ``liquidity pool attack", since 08 May 2022, \cite{LFG2022} (LFG) started to trade its reserves to restore the USD peg\footnote{According to \cite{LFG2022}, on 07 May 2022, the reserves included $80\,394$ BTC, $39\,914$ BNB, $6\,281\,671$ USDT, $23\,555\,590$ USDC, $1\,973\,554$ AVAX, $697\,344$ UST, $1\,691\,261$ LUNA.}. As a consequence of this, they were able to obtain some stability around $\$0.995$ until 09 May 2022. On this day, due to the strong and iterated selling pressure, UST lost its peg to USD for the second and last time. On 10 May 2022, LFG sold additional reserves to defend the peg, without success (see \citealp{LFG2022})\footnote{On 16 May 2022, the reserves included  $313$ BTC, $39\,914$ BNB, $1\,973\,554$ AVAX, $1\,847\,079\,725$ UST, $222\,713\,007$ LUNA.}. Terra reserves included a large amount of BTC, thus, according to social media rumors, attackers could have forced LFG to sell BTC on the market, decreasing its price, and increasing the attackers' profit, whose strategy relied on short selling positions against BTC (\citealp{Locke2022}). To save the Terra project, on 11 May 2022 10:10am (UTC), Terra Team announced the endorsement of the community proposal 1164 (\citealp{Kwon2022}), allowing a more efficient mechanism to burn UST and mint LUNA\footnote{According to \cite{Kwon2022}, this change should have allowed the system to absorb UST more quickly, reducing its supply and recovering the parity with USD, at the cost of lower prices for LUNA.}. Regardless of this, community's trust on the project was already lost, giving rise to the complete collapse of both tokens. LUNA and UST dropped from a value of $\$87$ and $\$1$, on 05 May 2022, to less than $\$0.00005$ and $\$0.2$, on 13 May 2022, respectively. Profits for potential attackers are estimated to be over 800 millions (\citealp{Locke2022}). On this point, it is worth noting that the ``coordinated attack", using BTC to destabilise UST peg, might not have happened at all, since claims of its existence are based on contradictory social media rumors (\citealp{Castillo2022}). In other words, the attack against Terra, which started with a ``liquidity pool attack" on Curve-3pool, and the potential existence of short selling positions against BTC, could have been two unrelated operations executed by different market actors.

The failure of Terra project represents an additional proof of the intrinsic fragility of algorithmic stablecoins which became already evident after the first large-scale crypto ``bank run" suffered by Iron Finance in June 2021. Also Iron Finance was a two-coin system, based on an algorithmic stablecoin (IRON) and on a token counterpart (TITAN), which collapsed after the cascade selloff of TITAN due to a significant selling pressure by some ``whales" (\citealp{IronFinance2021}). Indeed, the strategy used to bring down Iron Finance was very similar to the one potentially observable in the LUNA-UST case. More specifically, Iron Finance suffered from a ``liquidity pool attack", losing the peg to USD for a short time. Hours later, unknown market actors increased the selling pressure to spread panic, giving rise to the collapse of both IRON and TITAN (\citealp{Finematics2021}). Both projects (Iron Finance and Terra) shared a very similar blockchain framework, being prone to the same type of attacks. This duality shows that algorithmic stablecoins are prone to failure due to two main reasons. First, as stated by \cite{Clements2021}, they are \textit{built on the fragile foundation of relying on uncertain historical variables: they need a support level of baseline demand, they need participation of willing arbitrageurs, and they need an environment of informational efficiency. None of these factors are certain, and all of them have proven to be highly tenuous in the context of financial crises or periods of extreme volatility}. Second, to survive a crash, algorithmic stablecoins need an intrinsic worth, i.e. there should be ``genuine economic transactions" reliably performed with their tokens (\citealp{Calcaterra2020}). This was not the case of Terra, whose stablecoin was mainly used to obtain high interests through the Anchor protocol. On the contrary, history demonstrates that traditional fiat currencies are generally able to survive similar kind of attacks (e.g. Sterling attack in 1992, see \citealp{Fratianni1996}), since they have an intrinsic value as mean of exchange or storing value.

As an additional cause of reflection, we remark how, nowadays, the role of stablecoins has been only marginally studied by the scientific community. There are only few studies on stablecoins' stability (\citealp{Grobys2021}), their role in portfolio diversification (\citealp{Wang2020, Baur2021})  and crypto asset price formation (\citealp{Barucci2022a, Kristoufek2022}). To the best of our knowledge, the Terra-Luna case has not been addressed in the cryptocurrency literature. In this paper, we provide the scientific community with the first insights on the Terra project's failure by analysing hourly prices and transaction data through the instruments of network science and herding analysis.

\section{Data and quantitative nature of the events}\label{data}

\subsection{Hourly data analysis} \label{OHLCV_analysis}

We use hourly closing price data from the CryptoCompare database (\citealp{CC2022}) for 61 cryptocurrencies from 01 May 2022 00:00 to 16 May 2022 23:00 (UTC)\footnote{The list of cryptocurrencies used in this paper and their corresponding sector according to taxonomy by \citealp{Messari}, is provided as supplementary material.}. This widely used proprietary data source excludes exchanges containing spurious volumes (see \citealp{Alexander2020} and \citealp{VidalTomas2022a}), ensuring a high standard of data quality. Specifically, we use data from the Kraken digital currency exchange (the $4^{th}$ largest exchange for traded volume according to \citealp{Exchanges_Cap}). 

\begin{figure}[H]
	\centering	
	\caption{Rescaled hourly closing prices for LUNA, UST and BTC. Main trigger events are plotted with dotted lines: (a) 05 May 2022 12:00, (b) 07 May 2022 22:00, (c) 09 May 2022 14:00, (d) 11 May 2022 10:00.}
	\begin{center}
		\includegraphics[scale=0.45]{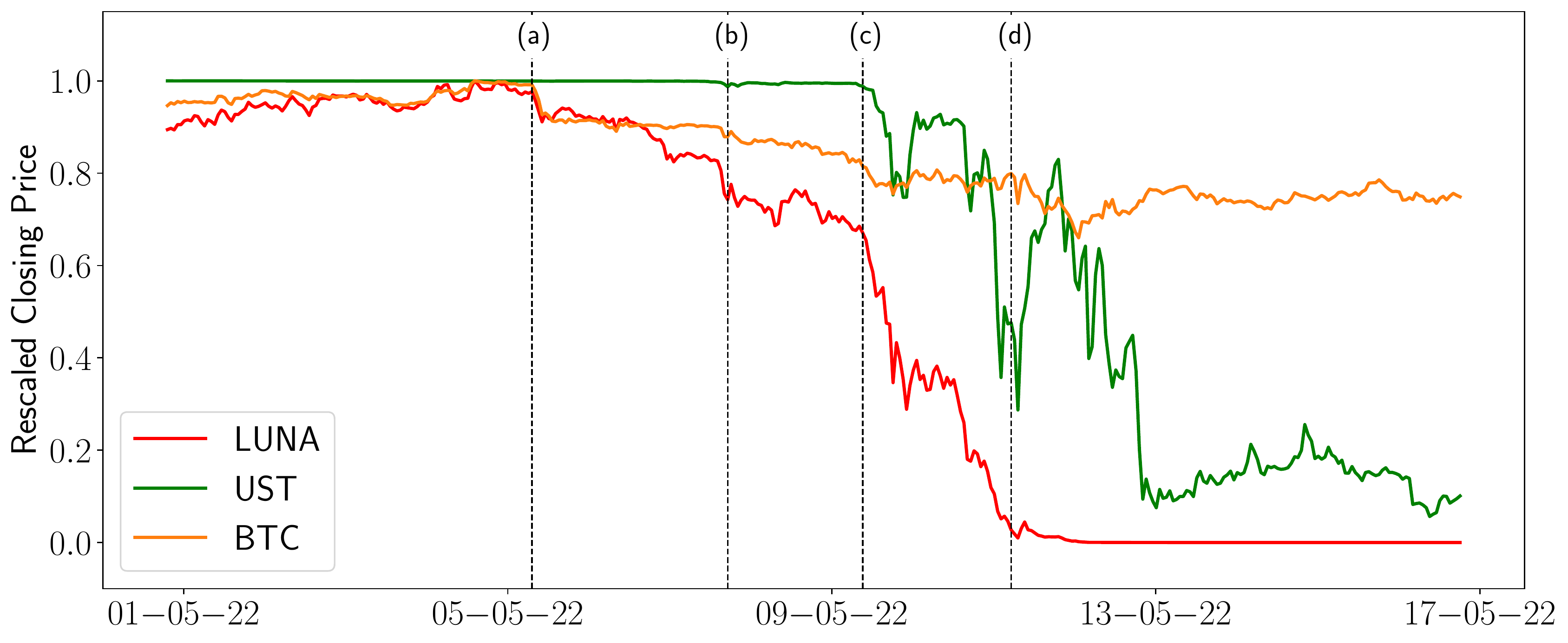}
	\end{center}
	\label{fig_des0}
\end{figure}

Figure \ref{fig_des0} shows rescaled hourly closing prices for LUNA, UST and BTC. Dotted lines underline trigger events that led to the collapse: (a) 05 May 2022 12:00, (b) 07 May 2022 22:00, (c) 09 May 2022 14:00, (d) 11 May 2022 10:00. Since (a) 05 May 2022 12:00, a strong and persistent selling pressure was evident on BTC and LUNA, with negative hourly log-returns\footnote{Hourly log-return $r$ for crypto asset $i$ at time $t$ is defined as $r_{i,t} = \log(p_{t})/\log(p_{t-1})$, where $p$ is the closing price.} at 13:00, 14:00 and 15:00, as reported in Table \ref{table_a}. On (b) 07 May 2022 22:00, UST lost its peg to USD for the first time as a consequence of the ``liquidity pool attack" on Curve-3pool (\citealp{Ashmore2022}; \citealp{Chainanalysis2022}). In response to this event, the LFG intervened, defending the UST peg and recovering the price to around $\$0.995$. However, on (c) 09 May 2022 14:00, UST lost its peg for the second (and last) time, giving rise to the main decrease in both LUNA and UST prices. Finally, on (d) 11 May 10:00, \cite{Kwon2022} presented the last attempt to defend the peg endorsing the community proposal 1164. After a beneficial impact that led the UST price to $0.8$\$, this announcement was interpreted by the market as the signal for the expected death of the Terra ecosystem, giving rise to the final crash in the market.

\begin{table}[H]
	\centering
	\small
	\renewcommand{\arraystretch}{0.7}
	\caption{Hourly log-returns for LUNA and BTC after 05 May 2022 12:00 (a).}
	\begin{tabular}{lcc}
		Date  & LUNA  & BTC  \\
		\midrule
		05/05/2022 13:00 & -0.0195 & -0.0133  \\
		05/05/2022 14:00 & -0.0240 & -0.0205 \\
		05/05/2022 15:00 & -0.0258 & -0.0339  \\
		\bottomrule
	\end{tabular}%
	\label{table_a}%
\end{table}%

In order to analyse dependency structures in cryptocurrency market during Terra collapse, for each cryptocurrency $i$, we compute hourly log-returns $r_{i,t}$. Moreover, to analyse herding behaviour, we also compute equally-weighted (EW) market returns ($r_{m,t}$). Table \ref{descriptive} reports descriptive statistics for LUNA, UST, BTC and the EW market, which shows the negative tendency of the market during this period. 

\begin{table}[H]
	\centering
	\small
	\renewcommand{\arraystretch}{0.7}
	\caption{Descriptive statistics of hourly log-returns, from 01 May 2022 00:00 to 16 May 2022 23:00 (UTC).}
    \begin{tabular}{lccccccc}   & Mean  & Median & Std   & Skewness & Kurtosis & Min.  & Max. \\
	\midrule
	LUNA & -0.03353 & -0.00264 & 0.32555 & -8.94240 & 133.56931 & -4.89285 & 1.13664 \\
	UST & -0.00543 & 0.00000 & 0.09613 & -1.60611 & 21.06592 & -0.76157 & 0.49813 \\
	BTC & -0.00061 & -0.00055 & 0.01133 & -0.05591 & 11.89428 & -0.07450 & 0.06266 \\
	\midrule
	(EW) Market & -0.00179 & -0.00130 & 0.02048 & -0.75148 & 18.46993 & -0.16394 & 0.13608 \\
	\bottomrule
\end{tabular}%
	\label{descriptive}%
\end{table}%

\subsection{Transaction data analysis}
\begin{figure}[H]
	\centering	
	\begin{center}
		\includegraphics[scale=0.31]{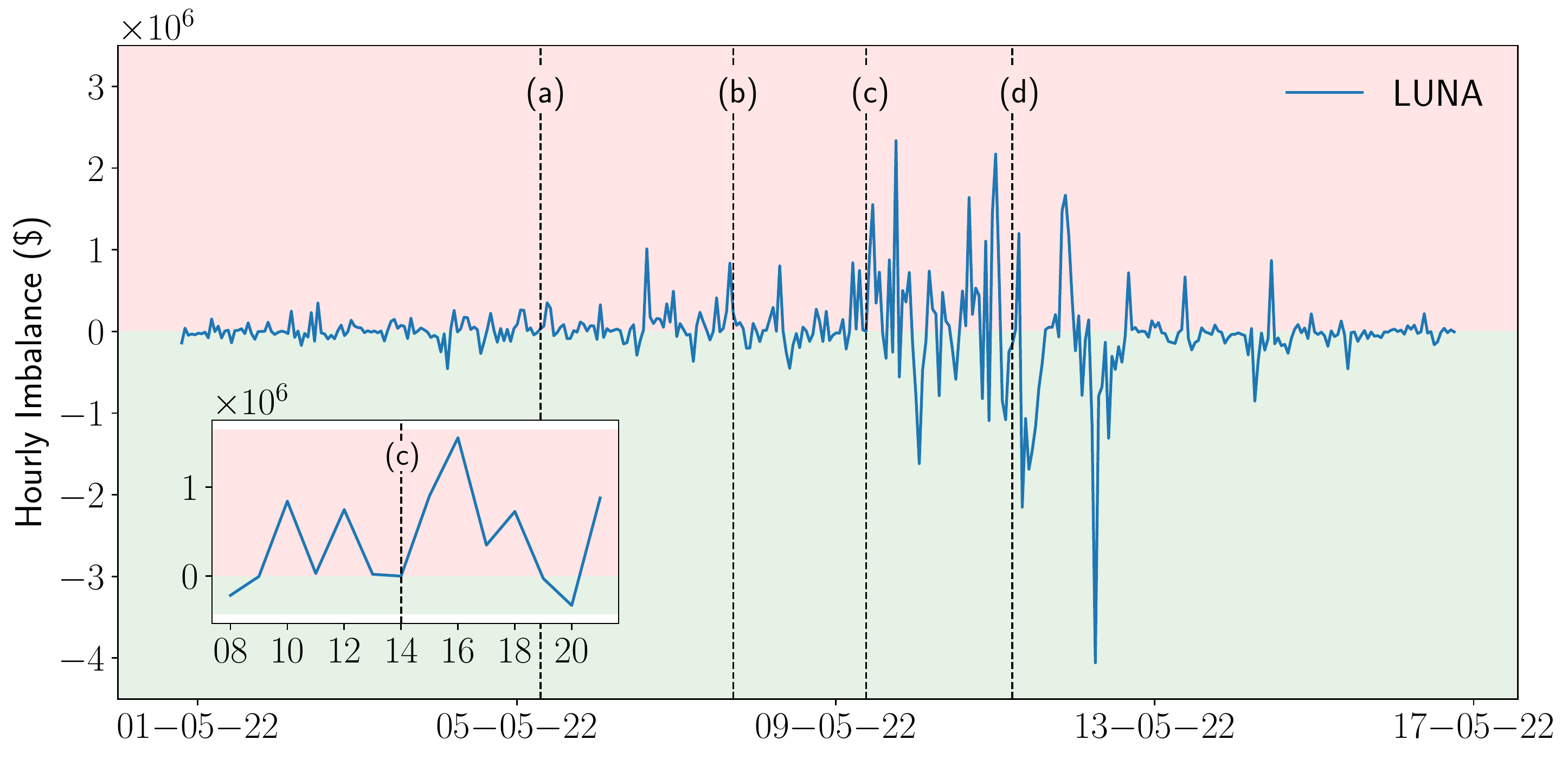}
		\includegraphics[scale=0.31]{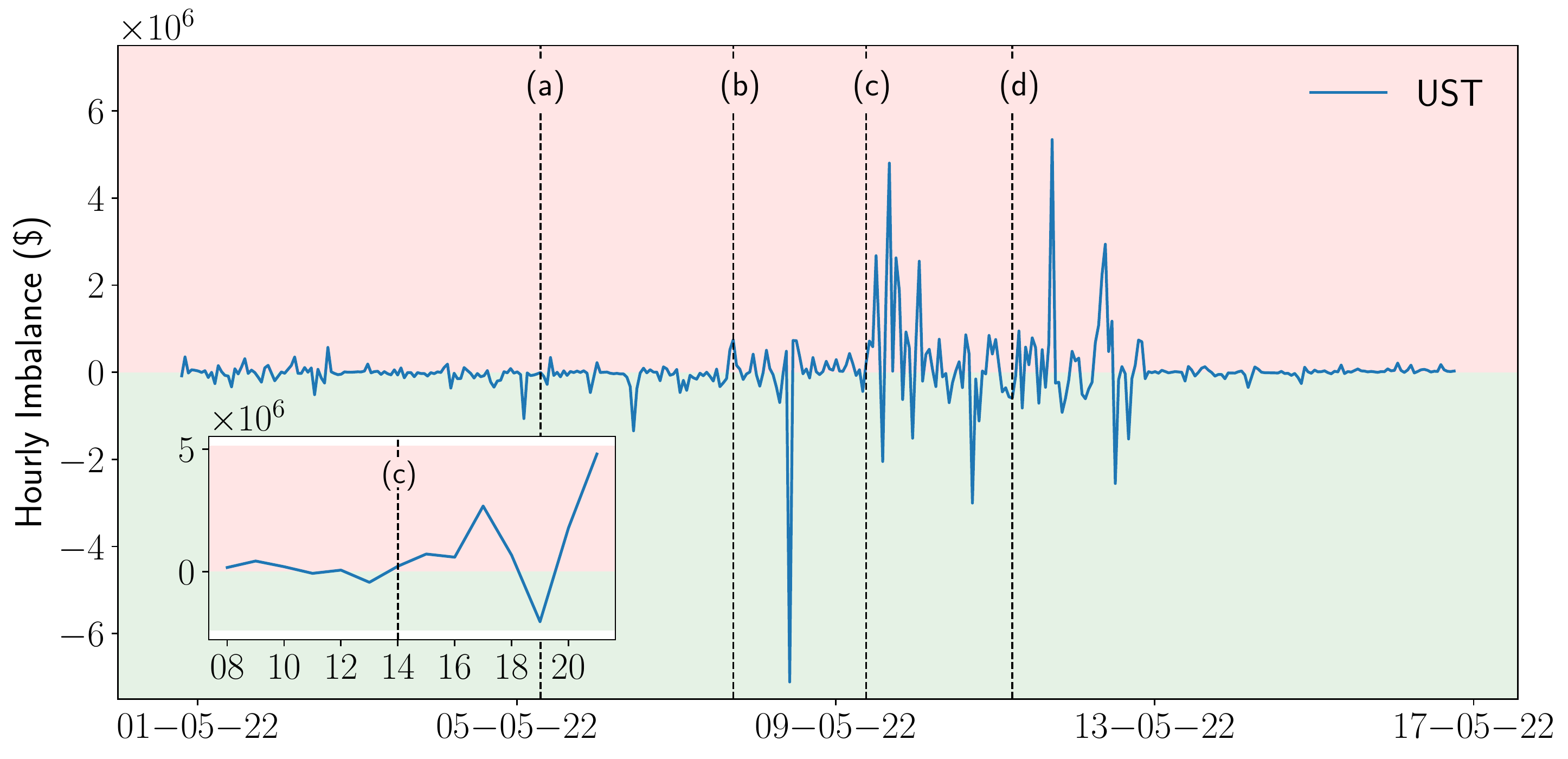}
		\includegraphics[scale=0.32]{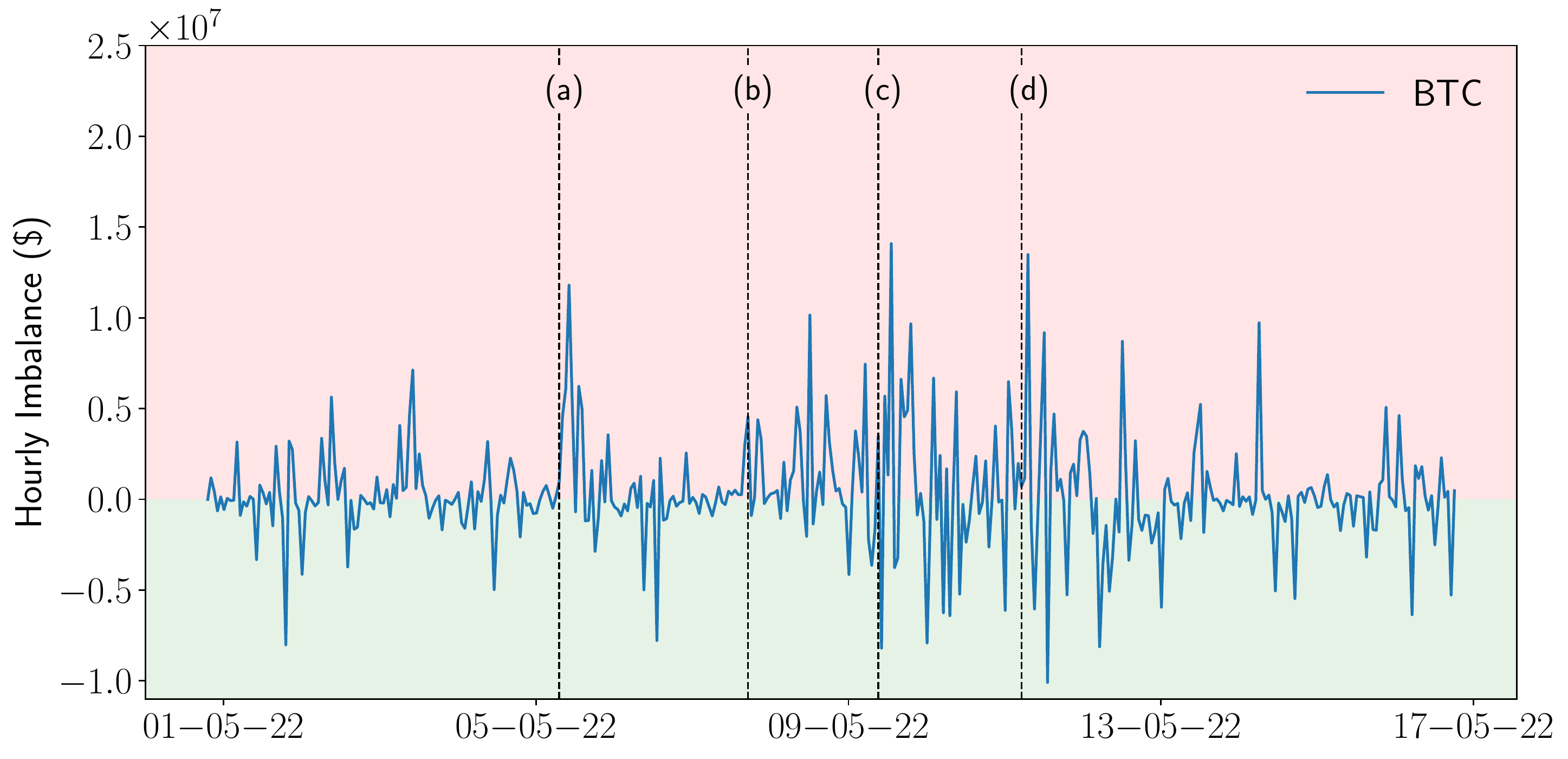}

	\end{center}
	\caption{Hourly imbalance for LUNA, UST and BTC in the Kraken digital currency exchange. Positive values (red color area) and negative values (green color area) denote selling and buying pressure, respectively.}
	\label{imbalance}
\end{figure}

To provide a complete overview of Terra project's collapse, we analyse public trades (i.e. transaction data) for BTC, LUNA and UST. Compared to hourly prices, which are retrieved from CryptoCompare database (\citealp{CC2022}), this new dataset is directly sourced from Kraken digital currency exchange using CCXT (\citealp{Ccxt2022}) Python package. Figure \ref{imbalance} reports hourly imbalances: positive values denote a selling pressure, while negative ones denote a buying pressure.\footnote{In order to compute the imbalance, we firstly distinguish between public trades on the buy and sell side. We then multiply the volume of each transaction and the price at which it is executed, obtaining the transaction's costs. We aggregate resulting transactions' costs by hour. We finally subtract buy hourly transactions costs from sell hourly transactions costs.} We observe interesting features that could be related to some of the events introduced in Section \ref{sec:Introduction}. First, on 05 May 2022 (a), we identify a strong, positive imbalance for BTC, with a value comparable to those detected during the main collapse of LUNA and UST (i.e. from 09 to 11 May 2022 (c-d)). We do not observe comparable high positive imbalance values for LUNA and UST on the same date. Therefore, even though we cannot confirm that short selling positions were opened against BTC as described by different social media sources (\citealp{Hall2022}; \citealp{Ashmore2022}; \citealp{Locke2022}), we remark the presence of a considerable selling pressure on this cryptoasset. In other words, if short selling positions were truly opened against BTC, 05 May 2022 should be the most plausible day in which this happened.\footnote{In this line, we also observe higher BTC selling pressure, from (b) to (c). Thus, on these days, market actors could have spread more panic in the market by selling BTC.} Second, focusing on the UST behaviour, we underline the existence of a remarkable selling pressure since 09 May 2022 (i.e. after the second UST de-pegging). This finding is particularly relevant, since the selling pressure after the first UST de-pegging (b) was remarkably lower than the second one, which could be in line with social media news when contending that attackers caused panic by selling whale-size UST holdings (\citealp{Hall2022}; \citealp{Ashmore2022}; \citealp{Locke2022}). This strategy would be similar to the one used to bring down Iron Finance, given that attackers waited to the second de-pegging, taking advantage of the uncertainty that already dominated the market (\citealp{Finematics2021}). Last but not least, it is relevant to emphasise that we cannot confirm the existence of a coordinated attack executed by related market agents. Indeed, the potential short selling against BTC and the Terra attack could have occurred at the same time by chance, in the context of a global economic uncertainty. In the case of Iron Finance collapse, no coordinated attack using a third cryptoasset was detected (\citealp{Finematics2021}).

\subsection{Anchor protocol}

\begin{figure}[H]
	\centering	
	\caption{Minimal schema representing the Anchor protocol's working mechanism. Users from Terra ecosystem were allowed to use Anchor protocol to (i) lend capital (UST) and (ii) borrow capital (UST) using bonded assets (bAssets) as collateral.}
	\begin{center}
		\includegraphics[scale=0.2]{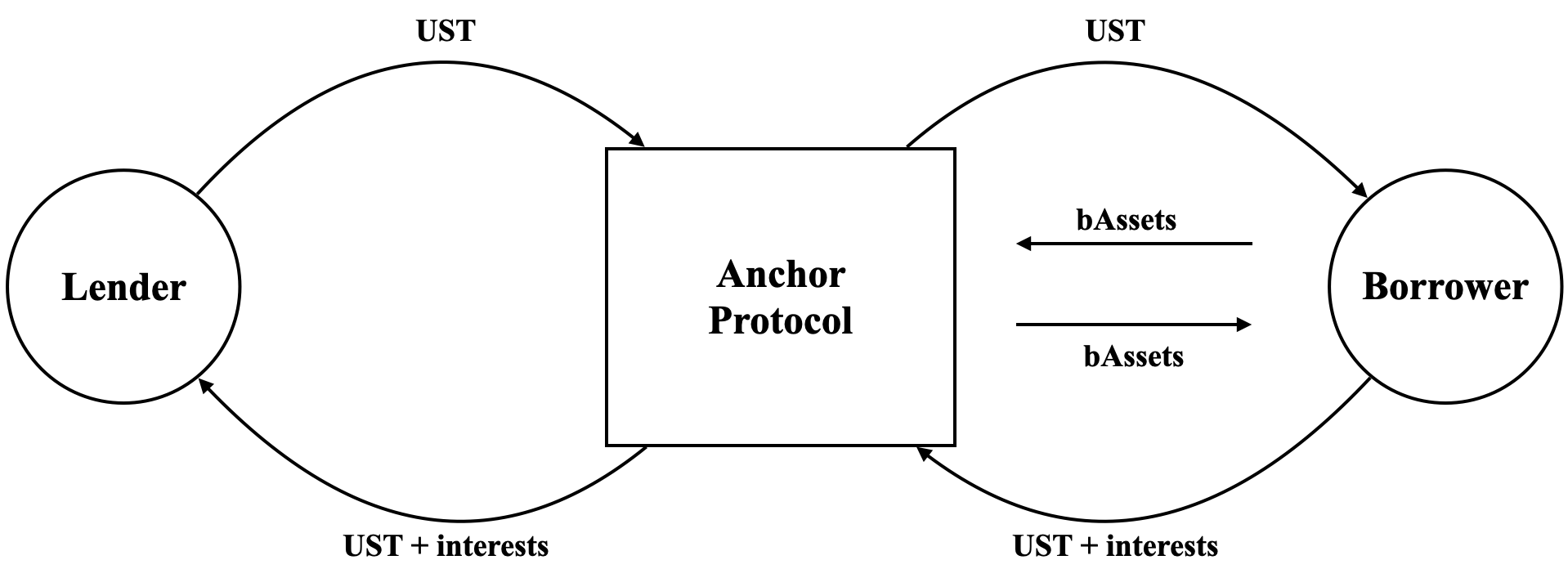}
	\end{center}
	\label{anchor_protocol_schema}
\end{figure}

Anchor (\citealp{Anchor2022}) was a savings protocol designed to allow users from Terra ecosystem to (i) lend capital (UST) and (ii) borrow capital (UST) using bonded assets (bAssets, i.e. tokens representing ownership of an asset on a Proof-of-Stake blockchain) as collateral (see Figure \ref{anchor_protocol_schema}). Anchor protocol was at the base of the fast expansion of Terra project since it offered to its depositors up to 20\% APY. This scheme was similar to the one used by Iron Finance, which offered more than 100\% APY (\citealp{IronFinance2021a}). 

\begin{table}[H]
	\centering
		\renewcommand{\arraystretch}{0.75}
	\caption{Main statistics for the Anchor protocol. Total deposit, borrow, and collateral are expressed in UST and millions. Market cap and circulating supply is also expressed in millions. The utilization ratio is defined as borrowed deposits divided by total deposits.}
		\begin{adjustbox}{max width=\textwidth}
    \begin{tabular}{lccccccccccc}
	& 17/06/2021 & 01/01/2022 & 01/05/2022 & 06/05/2022 & 08/05/2022 & 09/05/2022 & 10/05/2022 & 11/05/2022 & 12/05/2022 & 13/05/2022 & 14/05/2022 \\
	\midrule
	Total deposits & 358.93 & 5135.89 & 13714.53 & 14090.77 & 11775.4 & 9829.35 & 6465.31 & 4398.84 & 2595.82 & 1765.17 & 1503.79 \\
	Borrowed deposits & 128.8 & 2040.18 & 2929.29 & 3011.6 & 2424.88 & 1993.14 & 970.02 & 255.75 & 159.46 & 140.06 & 78.48 \\
	Utilization ratio & 35.88\% & 39.72\% & 21.36\% & 21.37\% & 20.59\% & 20.28\% & 15.00\% & 5.81\% & 6.14\% & 7.93\% & 5.22\% \\
	Total collateral & 422.55 & 5576.74 & 5719.24 & 5525.53 & 4447.74 & 2464.61 & 1940.05 & 611.09 & 266.06 & 139.49 & 59.57 \\
	\midrule
	&       &       &       &       &       &       &       &       &       &       &  \\
	\midrule
	Market cap UST & 1904.08 & 10133.54 & 18562.62 & 18725.63 & 18644.84 & 14761.64 & 14029.07 & 10718.07 & 4698.47 & 1737.13 & 2215.84 \\
	Price UST & 1.00  & 1.00  & 1.00  & 1.00  & 1.00  & 0.79  & 0.80  & 0.80  & 0.41  & 0.15  & 0.20 \\
	Circulating Supply & 1904.08 & 10133.54 & 18562.62 & 18733.12 & 18712.21 & 18605.55 & 17538.52 & 13379.19 & 11498.96 & 11280.09 & 11282.28 \\
	\midrule
	Anchor deposit/ Supply & 18.85\% & 50.68\% & 73.88\% & 75.22\% & 62.93\% & 52.83\% & 36.86\% & 32.88\% & 22.57\% & 15.65\% & 13.33\% \\
	\bottomrule
\end{tabular}%
\end{adjustbox}
	\label{anchor}%
\end{table}%

Looking at Table \ref{anchor} it is possible to see how, from 17 June 2021 to 06 May 2022, total UST deposits in Anchor increased by 3826\%\footnote{Data from Anchor protocol dashboard is available since 17 June 2021. No data are available on 07 May 2022.}. Consequently, before the Terra collapse, this protocol kept the 75\% of all the UST circulating supply, leaving the remaining 25\% as a means of exchange. According to \cite{Platias2020}, Anchor's mechanism was originally designed to allow the borrower to pay less interest as the utilization ratio (borrowed deposits divided by total deposits) decreased, resulting in lower interests for the depositor. However, despite the decrease in the utilisation ratio, the deposit interest rate was kept around 20\% since 2021 (\citealp{Anchortweet}; \citealp{Jung2022}). In addition, the deposit interest rate was computed considering the yields of all bAssets used as collateral for borrowing the stablecoin, while its value did not increase significantly since January 2022. The unsustainability of the protocol is confirmed by capital infusions into Anchor's reserves in February 2022 (\citealp{Kelly2022}) and by recurrent requests from community members for the modification of the interest rates' mechanism (\citealp{Anchor2022a}). Compared to fiat currencies, UST was mainly used for speculative purposes, which implied that it was more prone to extreme market events. Looking at Table \ref{anchor}, we observe that, from 06 May 2022 to 11 May 2022, the protocol lost around 9.5 billions in UST, i.e. the 69\% of the deposits. Thus, once it became evident that both the Terra system and the high-interest rate mechanism were at risk, depositors sold their UST igniting the bank run.

\section{Methodology}\label{methodology}

\subsection{Network analysis: Triangulated Maximally Filtered Graph (TMFG)}\label{methodology_network}
To get more insights into Terra project's failure, we describe the evolution of dependency structures among cryptocurrencies during the crash. In order to do this, we use instruments provided by network science. Networks have been extensively used to study financial systems (\citealp{Mantegna1999, Aste2010, Briola2022}) modeling dependencies among assets through correlations. Different correlation measures capture different relationships among assets. We decide to use the Pearson correlation coefficient to model linear relations among cryptocurrencies. It is worth noting that, in a condition of stress of the underlying system, pure correlations might be affected by an excessive sensitiveness. In order to partially mitigate such an effect, we assign a structure of weights to observational events, giving more relevance to the last observations of a given window. Weighted correlations are found to be smoother and recovering faster from market's turbulence than their unweighted counterparts, helping also to discriminate more effectively genuine from spurious correlations. Following the definition in \cite{Pozzi2012}, we define the Pearson correlation coefficient weighted with exponential smoothing as follows:

\begin{equation}
\label{eq:exponentially_smoothed_corr_coef}
    \rho_{i,j}^w = \frac{\sum_{t=1}^{\Delta t} w_t (y_t^i - \overline{y}_i^w)(y_t^j - \overline{y}_j^w)}{\sqrt{\sum_{t=1}^{\Delta t} w_t (y_t^i - \overline{y}_i^w)^2} \sqrt{\sum_{t=1}^{\Delta t} w_t (y_t^j - \overline{y}_j^w)^2}} .
\end{equation}

where $w_t = w_0 e^{\frac{t-\Delta t}{\theta}}, \forall t \in \{1, 2, \dots, \Delta t\} \land \theta > 0$ represents an exponentially smoothed weight structure such that $\sum_{t=1}^{\Delta t} w_t = 1$ and  $\overline{y}_k^w = \sum_{t=1}^{\Delta t} w_t y_t^k$. $\Delta t$ corresponds to rolling windows made of 24 hours with steps of 1 hour each and $\theta$ is here set to $0.3$.

Such a definition can be used to build a range of networks representing dependency structures among cryptocurrencies (\citealp{Briola2022}). Here we use the state-of-the-art methodology, namely, Triangulated Maximally Filtered Graph (TMFG) (\citealp{Massara2017}). Such a filtering network comes with several advantages compared to its alternatives. It is able to capture meaningful interactions among multiple assets and it is characterised by topological constraints which help to regularise for probabilistic modeling (\citealp{aste2022topological}). As a measure of network centrality, we compute the eigenvector centrality, which allows us to measure the influence of a crypto-asset in the system. Intuitively, a cryptocurrency has an higher eigenvector centrality as long as it is connected to other relevant cryptocurrencies, which are also characterised by an high eigenvector centrality. In order to highlight relevant events and reduce the impact of secondary ones, also in this case, we compute an exponentially smoothed rolling average of the signal with a smoothing factor equals to $0.3$.\footnote{Results are consistent to different values of $\theta$.}

\subsection{Herding analysis}\label{methodology_herding}

In this study, we also use the approach proposed by \cite{ChangChengKhorana2000} to study the possible emergence of herd behaviour during Terra collapse. This method is based on the notion of herding towards market consensus in which ``herds are characterised by individuals who suppress their own beliefs and base their investment decisions solely on the collective actions of the market, even when they disagree with its predictions" (\citealp{ChristieHuang1995}). More specifically, \cite{ChangChengKhorana2000} analysed the existence of herding in a system of $N$ assets through the cross-sectional absolute deviation of returns (CSAD) as a measure of return dispersion,

\begin{equation}
CSAD_{m,t}=\frac{\sum_{i=1}^{N} |r_{i,t}-r_{m,t}|}{N},
\end{equation}
and regressing the CSAD of returns with respect to market returns,

\begin{equation}
CSAD_{m,t}=\alpha+\beta_{1} |r_{m,t}|+\beta_{2} r_{m,t}^{2}+u_t
\label{CSAD1}
\end{equation}
where $|r_{m,t}|$ is the absolute term and $r_{m,t}^{2}$ denotes the square of market returns. \cite{ChangChengKhorana2000} noted that rational asset-pricing models imply a linear relation between the dispersion in individual asset returns and the return on a market portfolio. As the market’s absolute return grows, cross-sectional return dispersion would also be expected to rise, in a linear fashion. Conversely, during periods of euphoria or fear, investors may react in a more uniform manner, giving rise to herding behavior. This phenomenon will decrease the dispersion among returns or at least increase at a less-than-proportional rate with the market return. Therefore,  the emergence of herding behaviour will be consistent with a significantly negative coefficient $\beta_{2}$ in Eq.(\ref{CSAD1}). However, in the absence of herding, we should only observe a positive and linear relation between the cross-sectional return dispersion and absolute market returns, i.e. a significant positive coefficient $\beta_{1}$, and a non-significant coefficient $\beta_{2}$. In this line, given that Eq.(\ref{CSAD1}) represents the symmetric form that includes all the market return distribution, following \cite{Cui2019} and \cite{Chiang2010} we also divided market returns to distinguish asymmetric herding when the market is up or down,
\begin{equation}
CSAD_{m,t}=\alpha+\beta_{1} (1-D)|r_{m,t}|+\beta_{2}D |r_{m,t}|+\beta_{3} (1-D)r_{m,t}^{2}+\beta_{4}D r_{m,t}^{2}+u_t
\label{CSAD2}
\end{equation}
where $(1-D)$ and $D$ are dummy variables equal to 1 when $r_{m,t}\geq 0$ and $r_{m,t}< 0$, respectively.\footnote{Following the corresponding literature (e.g. \citealp{ChangChengKhorana2000}; \citealp{Chiang2010}), in Eq.(\ref{CSAD1}) and Eq.(\ref{CSAD2}), we used Newey-West standard errors (\citealp{Newey1994}).}

\section{Results}\label{results}

\subsection{Correlations and network analysis}\label{results_network}

\begin{figure}[H]
	\centering	
	\begin{center}
		\includegraphics[scale=0.3]{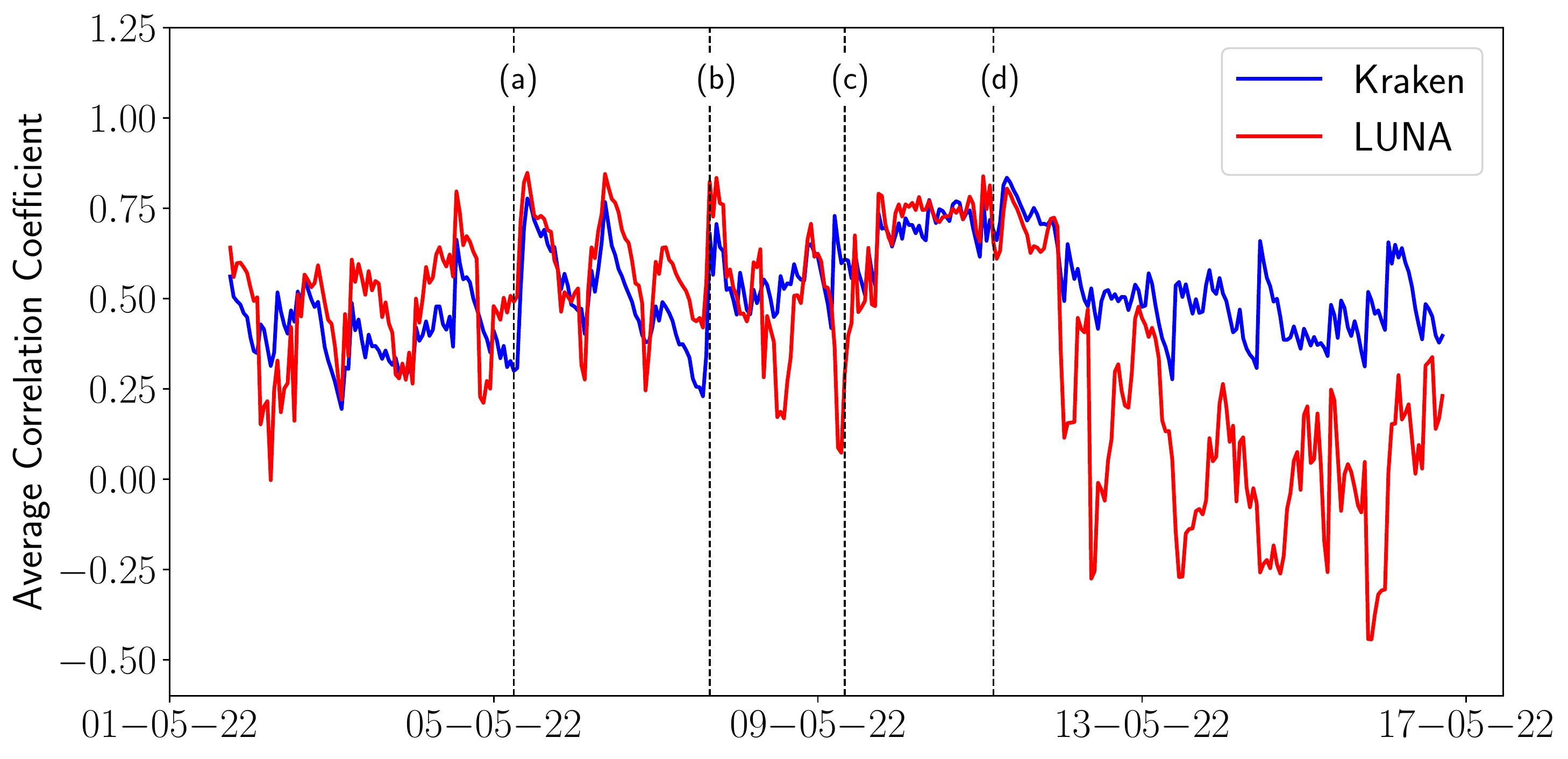}
		\includegraphics[scale=0.3]{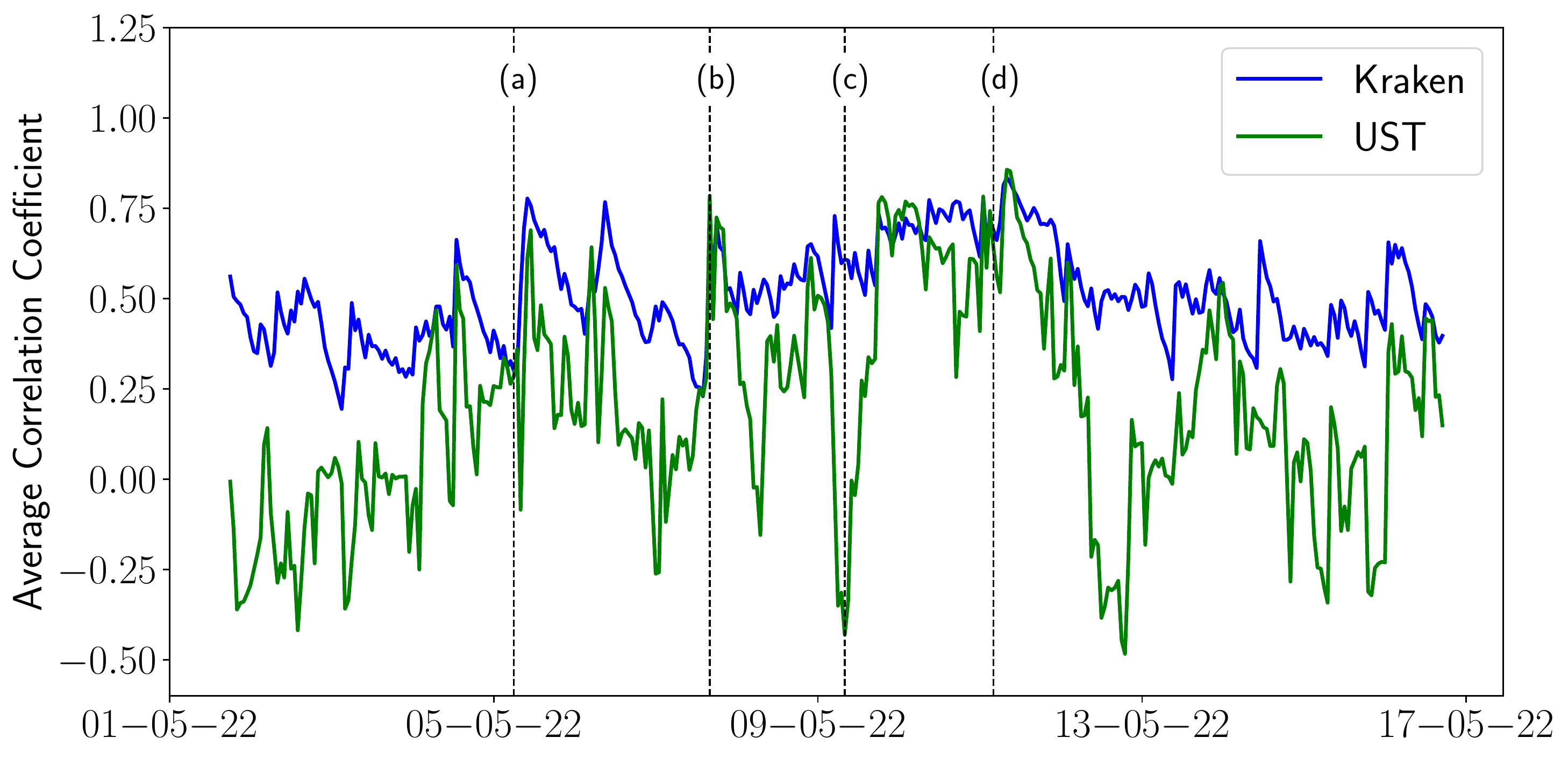}
		\includegraphics[scale=0.3]{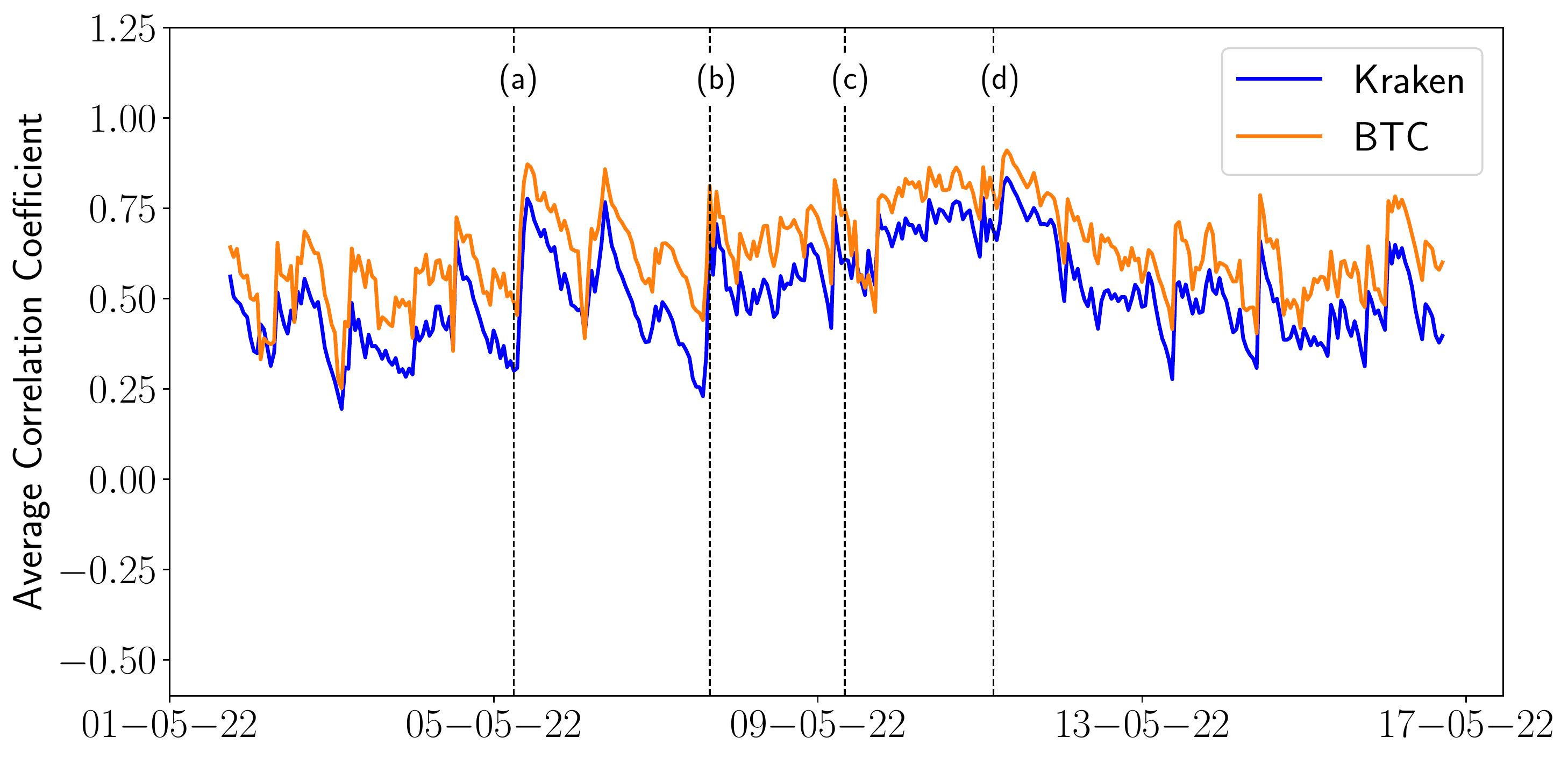}
	\end{center}
	\caption{Exponentially smoothed weighted correlations for LUNA, UST and Kraken digital currency exchange, using 24 hour rolling windows with steps of 1 hour each. Kraken (blue line) denotes the average correlation of all the 61 cryptocurrencies, while LUNA (red line), UST (green line) and BTC (orange line) represent their average correlation with the rest of the system.}
	\label{netfig1}
\end{figure}

Figure \ref{netfig1} reports exponentially smoothed average correlation coefficients for LUNA, UST, BTC and Kraken exchange\footnote{Kraken digital currency exchange represents the average correlation of all the 61 cryptocurrencies at our disposal.}. We observe that, until 09 May 2022 12:00 (c), the market (Kraken) showed sporadic peaks of high correlations, such as the one observed after the first UST de-pegging (b). Starting from the second UST de-pegging, between 09 May 2022 12:00 (c) and 11 May 2022 10:00 (d), we identify a continuous increase in market correlations, until it stabilised for a short time around $0.8$. Considering that we use exponentially smoothed weighted correlations, which weight more the last hours of the window, we state that this constant increase in correlations shows that the market kept a high co-movement at each hour until 11 May 2022.

Therefore, from (c) to (d), we detect the specific period when most of the market reacted to Terra collapse, coinciding with the main decrease in UST and LUNA prices (see Figure \ref{fig_des0}). Afterwards, we observe a decrease in correlations in LUNA, UST, BTC and Kraken digital currency exchange. This new scenario shows that the first two cryptocurrencies were ``excluded" from the system given their ``death", while the system seems to recover a lower and more stable degree of correlation.

\begin{figure}[H]
	\centering	
	\caption{Eigenvector centrality of LUNA, UST and BTC using 24h rolling windows. Color areas show the distribution of the eigenvector centrality for the rest of cryptocurrencies considering 1\%-99\%, 5\%-95\%, and 25\%-75\% percentiles.}
	\label{netfig2}
	\begin{center}
		\includegraphics[scale=0.3]{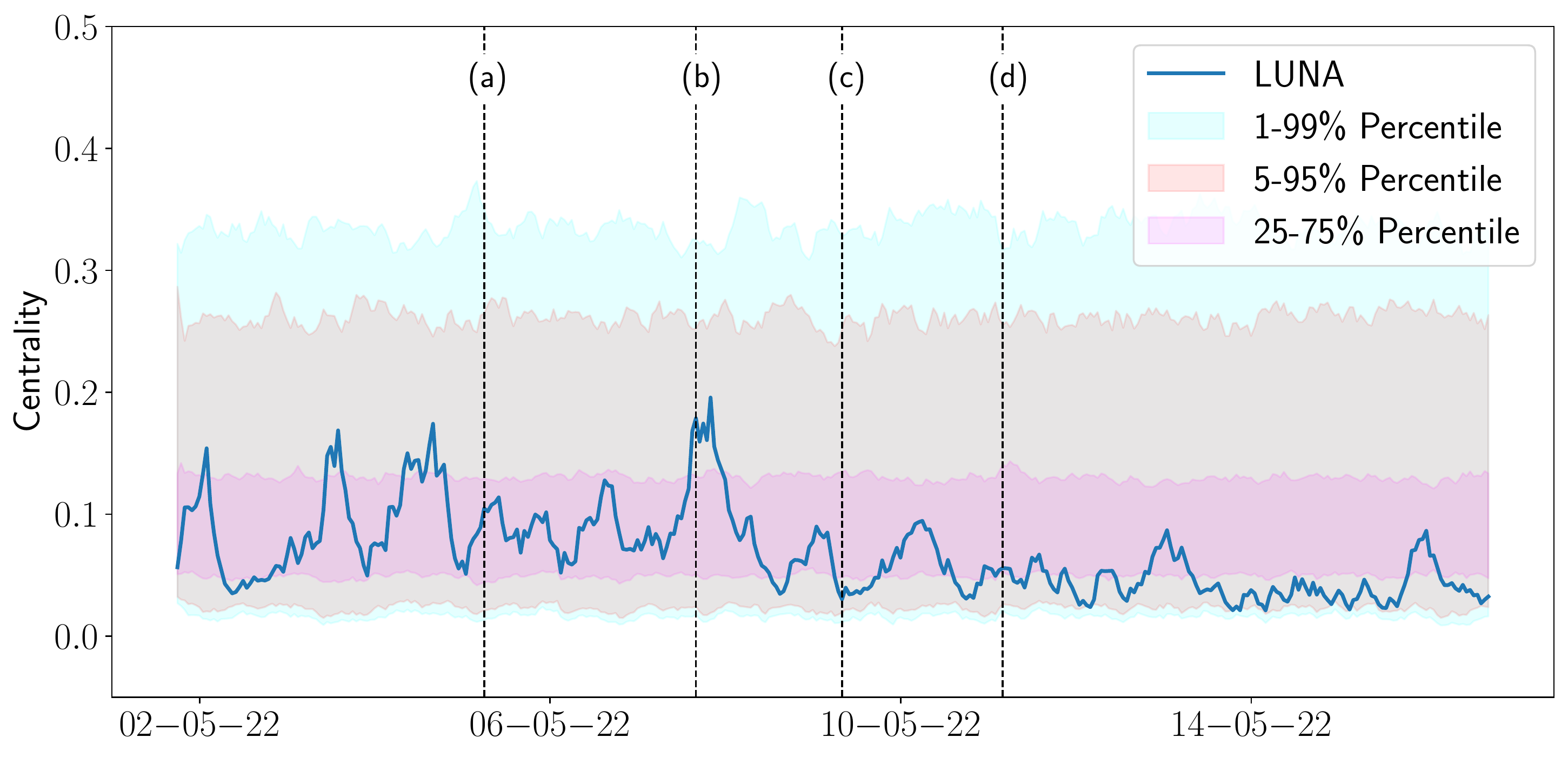}
		\includegraphics[scale=0.3]{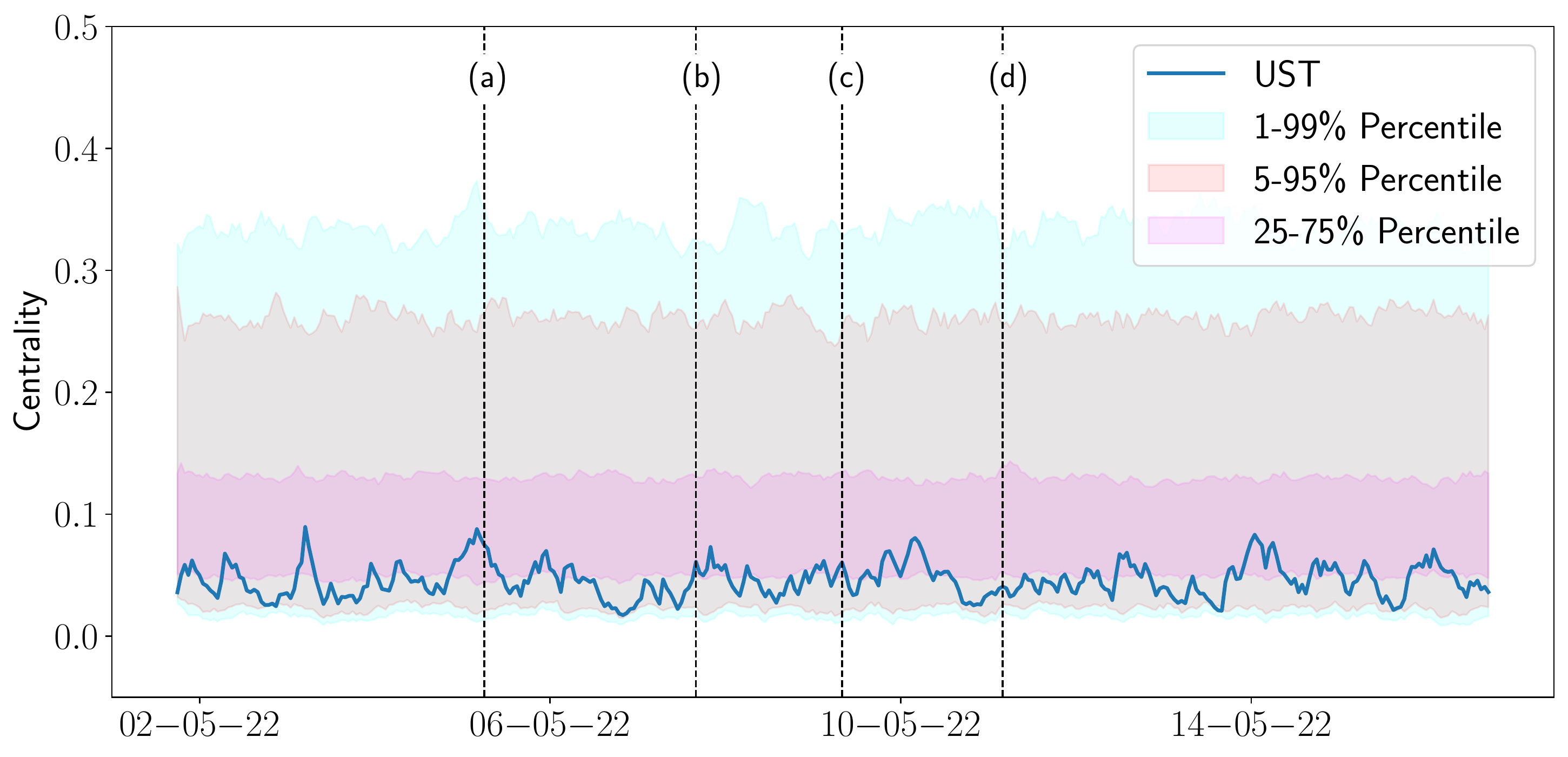}
		\includegraphics[scale=0.3]{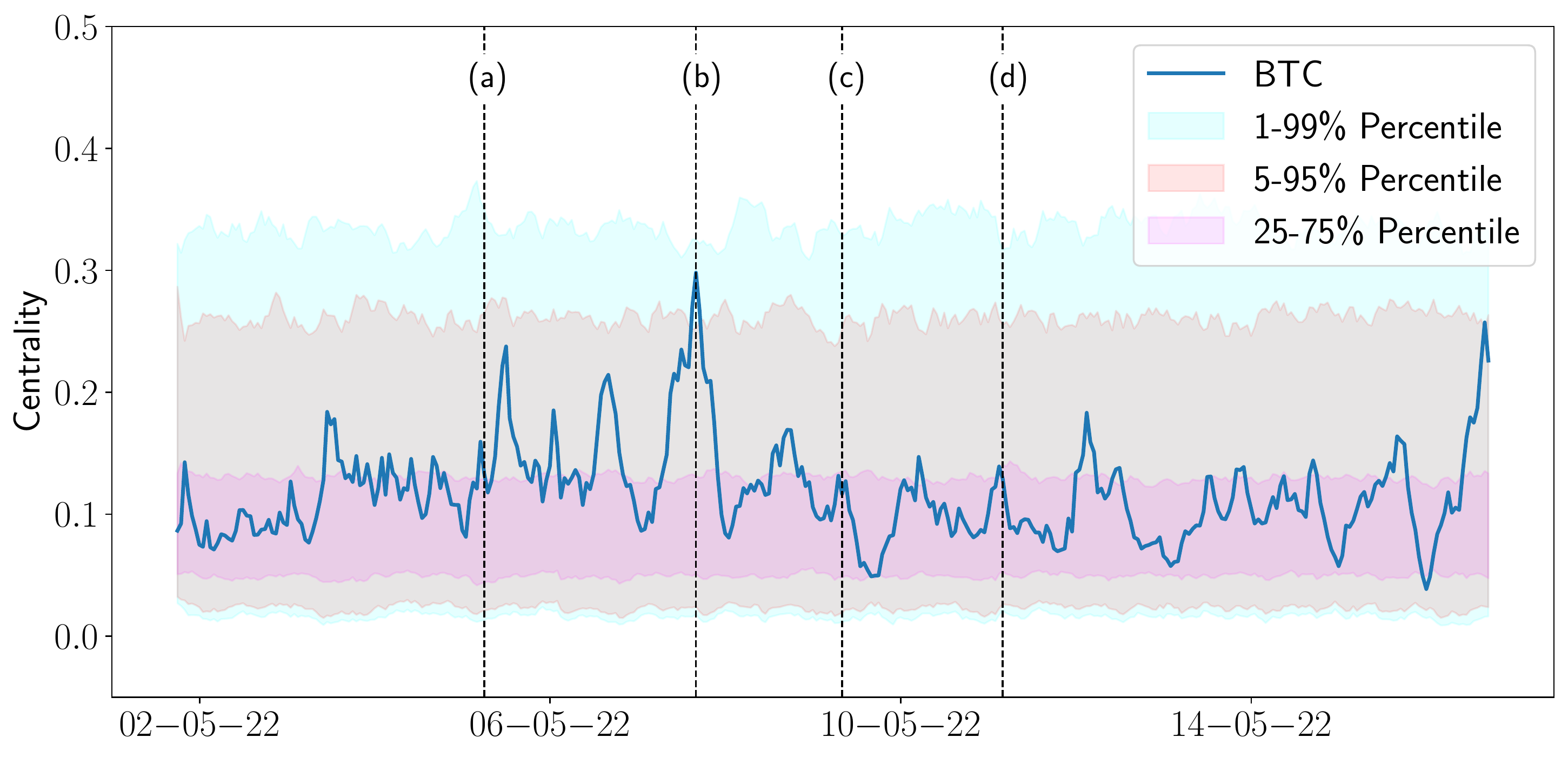}
	\end{center}
\end{figure}

\begin{figure}[h]
	\centering	
	\begin{center}
		\includegraphics[scale=0.21]{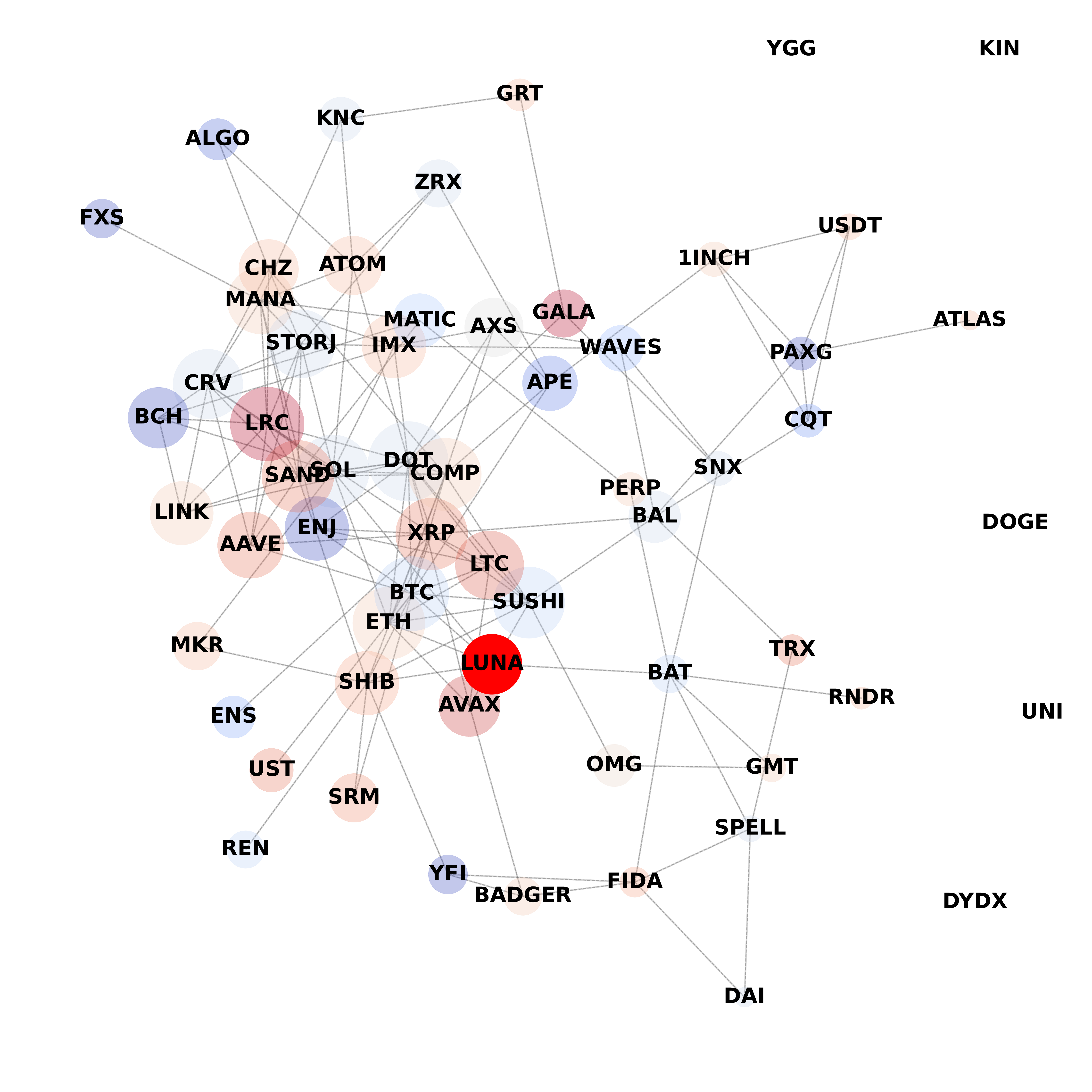}
	\end{center}
	\caption{TMFG on 7 May 2022 22:00 UTC (b). As an effect of the first Terra de-pegging, LUNA became central into cryptocurrency market's dependency structure, influencing it for a short time.}
	\label{net_b}
\end{figure}

\begin{figure}[h]
	\centering	
	\begin{center}
		\includegraphics[scale=0.545]{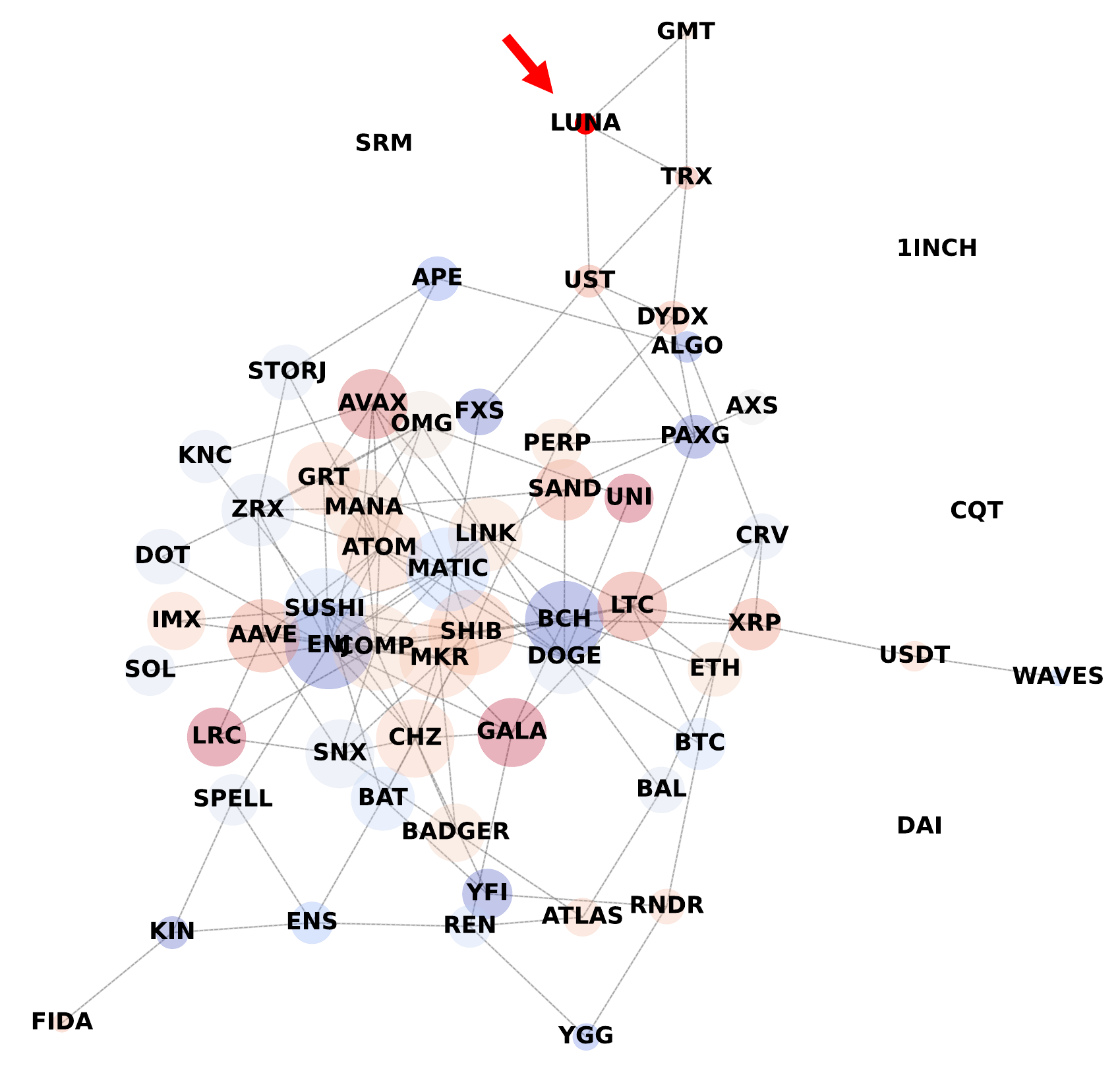}
	\end{center}
	\caption{TMFG on 9 May 2022 12:00 UTC (c). As an effect of the second Terra de-pegging, LUNA was completely excluded from the core of the network.}
	\label{net_c}
\end{figure}

After obtaining an initial picture through the study of market correlations, we compute the TMFG to obtain a better description of the system's collective dynamics. Figure \ref{netfig2} reports the evolution of TMFG's eigenvector centrality, through 24 hour rolling windows, for LUNA, UST and BTC. From (a) 05 May 2022 12:00 to (b) 07 May 2022 22:00, we observe that BTC was influencing the rest of the system, given its higher eigenvector centrality. This finding potentially supports the hypothesis of its involvement during the initial down-market since 05 May 2022 12:00. On (b) 07 May 2022 22:00, we identify the effect of the first Terra de-pegging on the cryptocurrency market's dependency structure, since LUNA became part of the core of the network, influencing the cryptocurrency market for a short time (see Figure \ref{net_b}). Since that day, LUNA was excluded from the core of the network, gradually lowering its eigenvector centrality. This finding can be observed in the network reported in Figure \ref{net_c}, during the second UST de-pegging on 09 May 2022 12:00 (c). Finally, after (d) 11 May 2022 10:00, LUNA was completely excluded from the system, being relegated in the lowest eigenvector percentile.\footnote{Regarding UST, we do not observe relevant changes in terms of eigenvector centrality given its nature as stablecoin. Thus, even though connections with other cryptocurrencies could exist, most of them were non-significant.}

\subsection{Herding analysis: CSAD approach}\label{results_herding}

Table (\ref{herd_table}) shows the results of the analysis of the herding behaviour in the cryptocurrency market during Terra collapse. Interestingly, despite the fact that the cryptocurrency market is more interconnected over time (\citealp{Aslanidis2021}) and herding is traditionally found on periods of uncertainty (\citealp{GamaSilva2019}), we do not observe herding during the Terra collapse. In particular, we report significant positive coefficients for $|r_{m,t}|$, $(1-D)|r_{m,t}|$, and $D |r_{m,t}|$, while we underline non-significant coefficients for $r_{m,t}^{2}$, $(1-D) r_{m,t}^{2}$ and $D r_{m,t}^{2}$.\footnote{For robustness purposes, we applied Eq.(\ref{CSAD2}) through rolling windows of 7 days, i.e. 168 hours/observations. However, the absence of herding remains. Moreover, if we remove stablecoins from our sample, whose returns are mainly 0 (e.g. USDT or DAI), we keep obtaining the same result.} This result supports the existence of an heterogeneous performance in the cryptocurrency market during this period. In Table \ref{returns_cd} we report average hourly returns from (b) 07 May 2022 22:00 to (d) 11 May 2022 10:00, where we observe that some cryptocurrencies, like BTC and ETH, were characterised by a ``better" (less negative) performance, compared to the rest of the market, whose average was equal to $-0.0031$. Consequently, even though we observe higher Pearson correlation coefficients during the collapse [see Figure \ref{netfig1}], UST-LUNA failure did not dramatically affected the market due to the absence of herding and the low reaction of BTC [see Figure \ref{net_b}]. We could conjecture that investors considered this event as a non-structural shock, which did not posed real risk to jeopardise the future of the cryptocurrency system.

\begin{table}[H]
	\centering
	\renewcommand{\arraystretch}{0.7}
	\small
	\caption{Regression results of $CSAD_{t}$ on market returns. Both generalised form and distinguishing between positive and negative market returns. Significance at the 1\%/5\%/10\% level is denoted by ***/**/*.}
			\begin{adjustbox}{max width=\textwidth}
	\begin{tabular}{cccc|cccccc}
		$\alpha$ & $|r_{m,t}|$  & $r_{m,t}^{2}$    & $\bar{R}^{2}$  & $\alpha$ & $(1-D)|r_{m,t}|$   & $D |r_{m,t}|$ & $(1-D)r_{m,t}^{2}$   & $D r_{m,t}^{2}$   & $\bar{R}^{2}$ \\
		\midrule
		0.0051*** & 0.5641*** & -0.4237 & \multirow{2}[2]{*}{0.49} & 0.0052*** & 0.5336*** & 0.5771*** & -0.0307 & -0.5851 & \multirow{2}[2]{*}{0.454} \\
		(0.0008) & (0.1040) & (0.5993) &       & (0.0008) & (0.0982) & (0.1323) & (0.7533) & (0.6861) &  \\
		\bottomrule
	\end{tabular}%
\end{adjustbox}
	\label{herd_table}%
\end{table}%

\begin{table}[htbp]
  \centering
  \small
 \caption{Average hourly log-returns from (b) 07 May 2022 22:00 to (d) 11/05/2022 10:00. First row of cryptocurrencies shows the most negative average returns, while the second row shows the most positive returns.}
    \begin{tabular}{ccccccccccc}
    Crypto (-) & LUNA  & UST   & SPELL & YGG   & FXS   & GMT   & APE   & ENS   & KNC   & ZRX \\
    \midrule
    Average returns & -0.0333 & -0.0088 & -0.0060 & -0.0057 & -0.0054 & -0.0054 & -0.0049 & -0.0043 & -0.0042 & -0.0042 \\
          &       &       &       &       &       &       &       &       &       &  \\
    Crypto (+) & YFI   & COMP  & BCH   & TRX   & BTC   & ETH   & PAXG  & USDT  & DAI   & MKR \\
    \midrule
    Average returns & -0.0013 & -0.0013 & -0.0013 & -0.0012 & -0.0012 & -0.0008 & -0.0002 & 0.0000 & 0.0000 & 0.0040 \\
    \end{tabular}%
  \label{returns_cd}%
\end{table}%

\section{Robustness analysis}
\begin{figure}[H]
	\centering	
	\begin{center}
		\includegraphics[scale=0.31]{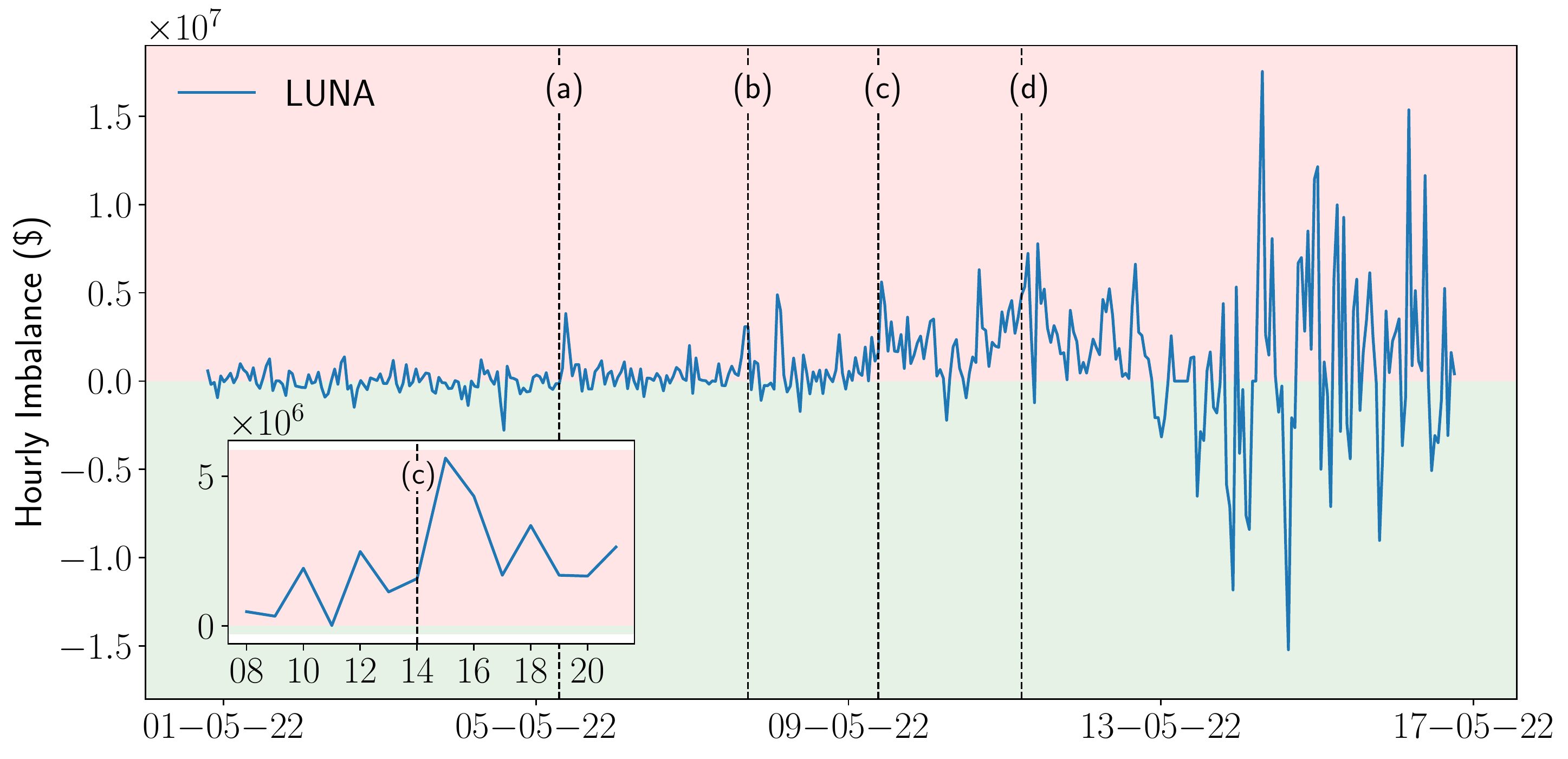}
		\includegraphics[scale=0.31]{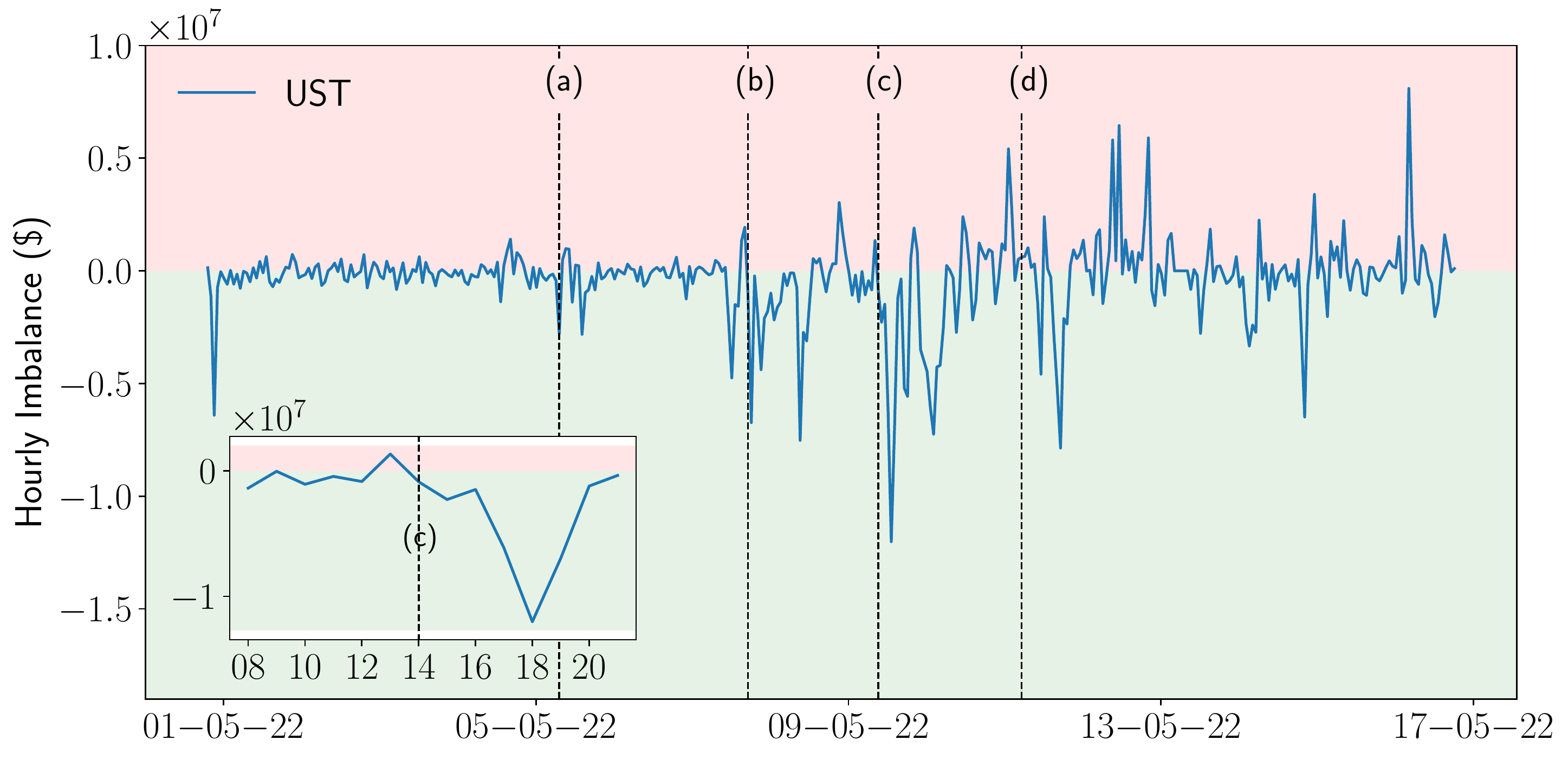}
		\includegraphics[scale=0.32]{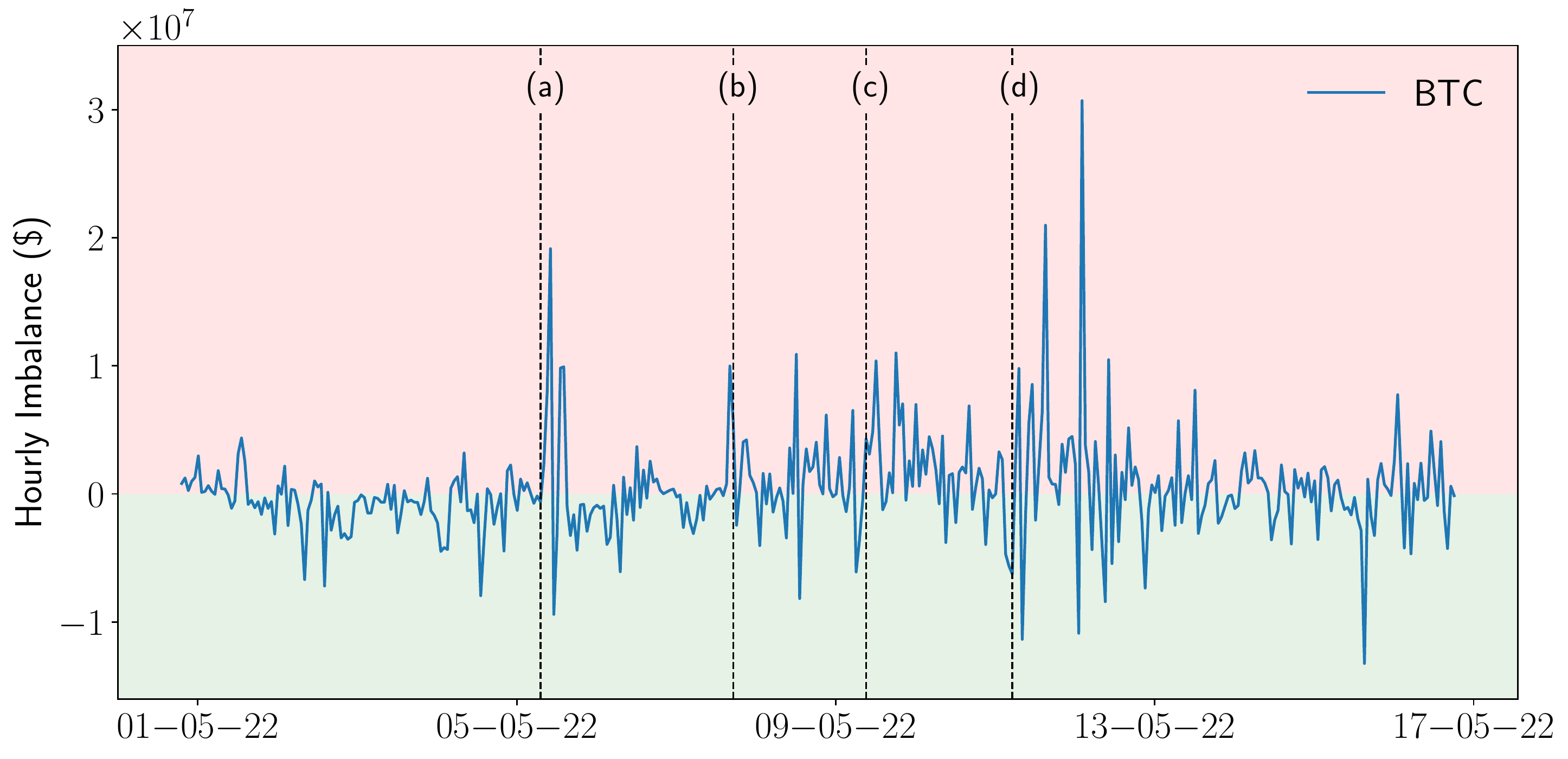}

	\end{center}
	\caption{Hourly imbalance for LUNA, UST and BTC. Positive values (red color area) and negative values (green color area) denote selling and buying pressure, respectively. Binance exchange.}
	\label{imbalance_binance}
\end{figure}

In this section, we use data from Binance digital currency exchange (the largest digital currency exchange platform according to \cite{Exchanges_Cap}) to assess the robustness of our results. Data retrieving is entirely accomplished using the CCXT (\citealp{Ccxt2022}) Python package. Compared to Kraken, Binance's cross-rates are expressed in BUSD (BinanceUSD), which could give rise to some inconsistencies in comparative studies (see e.g. \citealp{Alexander2020}). Consequently, we use it only in the context of robustness analysis. For the sake of space, we only report results where we observe more significant differences (see Figure \ref{imbalance_binance}). The rest of the results are consistent with the ones observed in Kraken and are available upon request. In Figure \ref{imbalance_binance} we show two relevant findings. First, in line with Kraken (see Figure \ref{imbalance}), on 05 May 2022 (a), we identify a high positive imbalance for BTC, which is remarkably lower only than a peak after the Terra project's collapse. This result enforces the finding that if short-selling positions were opened against BTC, this potentially happened on 05 May 2022. Second, focusing on UST, contrary to what observed on Kraken, we find a considerable degree of buying pressure during the Terra project's collapse. This finding could imply that Binance was used by LFG and/or other market actors to defend the UST peg.\footnote{In the case of LUNA, we observe high instability since 13 May 2022, which could be caused by speculators trading when the price was close to 0 or by the run to positions' liquidation.} 

\section{Implications and future research}\label{im_fr}
\subsection{Relevance for stakeholders}
Considering that the failure of large stablecoins built upon inappropriate decentralized finance frameworks (e.g. lack of collateral and dependencies with liquidity providers) could give rise to potential financial stability risks, we believe that our results are relevant for both investors and policymakers. Applications of stablecoins within the crypto space remarkably increased since 2020, while poorly supported by adequate regulation. Initially, they were used as: (i) bridge between fiat currencies and crypto-assets; (ii) ``parking space" for crypto volatility (\citealp{Adachi2022}). Nowadays, they are also used to provide most of the liquidity in DeFi applications such as decentralised exchanges and lending protocols (e.g. UST and Anchor protocol). Consequently, stablecoins play a critical role within the crypto space, where they are involved in almost 75\% of the entire trading (\citealp{Adachi2021}). Largest stablecoins represent a primary source of risk according to the European Central Bank (ECB), which states that \textit{a ``bank run" on Tether could disrupt trading and price discovery in crypto-asset markets, which could turn disorderly} (\citealp{DirectorateGeneralMacroprudentialPolicy2022}). Therefore, even though the collapse of UST was not able to rise relevant herding effects, the failure of larger stablecoins could generate systemic effects on the whole crypto universe. In addition, considering the growing interlinks between crypto-assets and the traditional financial system, the failure of stablecoins could also have implications external to the crypto space. These scenarios remark the need for a Global Stablecoin Standard and continuously updated regulations able to guarantee (i) a process for fulfilling holders' redemption claims, (ii) proper transparency of reserve asset composition and (iii) development of appropriate risk-management frameworks. Relevant regulatory standard-setting bodies (Financial Stability Board and Bank of International Settlements; see, e.g. \citealp{BIS2022}) and governments (Japan, Hong Kong, United Kingdom, European Union and the United States; see, e.g. \citealp{Asmakov2022}, \citealp{EC2022}) are already proposing robust rules for these digital assets. It would be beneficial that contributions from inside the DeFi community would also be proposed.

\subsection{Future lines of research}
Most of the existing literature on stablecoins concerns their stability (\citealp{Grobys2021}), their role in portfolio diversification (\citealp{Wang2020, Baur2021}), and crypto asset price formation (\citealp{Barucci2022a, Kristoufek2022}). Due to the nascent interest in this research field and the rapid growth of decentralised finance, gaps in the literature still exist. Recent events around Terra project highlights the need to analyse specific topics that have been neglected until now: (i) stablecoins as a means of exchange, (ii) stablecoins’ role within the crypto space, (iii) regulation on stablecoins and related protocols and (iv) potential risk to financial stability. First, stablecoins could not be appropriate as a mean of exchange due to their vulnerability to ``bank runs" and high transaction costs compared to non-blockchain alternatives (\citealp{Mizrach2022}). Second, the role of stablecoins as ``bridge" in the crypto space could be considered a double-edged sword since they could jeopardise the system's stability once becoming systemic. Third, the dependence of UST on Anchor protocol remarks the need for better scrutiny of third party protocols related to any stablecoin. Finally, it would be necessary to consider scenarios in which specific fragilities within the stablecoins' ecosystem may give rise to systemic financial stability risks.

\section{Conclusion}\label{conclusion}
In this paper, we review the Terra project's main features and describe the mechanisms that led to its failure. Our contribution to the existing literature is threefold. We first systematically organise news from heterogeneous sources to reconstruct a reliable timeline for the Terra project's collapse. We hence identify four main breakpoints for the crash and, starting from hourly data for 61 cryptocurrencies from Kraken digital currency exchange, we quantitatively characterise them. We exploit the power of smoothed weighted correlations to build state-of-the-art graph structures which demonstrate to be both able to efficiently capture dependency structures among cryptocurrencies and robust to extreme market conditions. As a further  element of novelty, we enforce our analysis using transaction data to detect relevant micro-structural market events.

Combining these last two approaches, we uncover BTC's reference role during the first phase of the collapse. In particular, we outline the existence of intense selling pressure on this crypto asset and identify 05 May 2022 as the potential fuse for the process that led to the Terra project's failure. We remark that it is impossible to conclude that this event and the Terra project's collapse were part of a coordinated strategy (\citealp{Castillo2022}). The weakness of the current global economy (e.g. Russia - Ukraine conflict (\citealp{Boungou2022}), bear markets in the main financial indices (\citealp{Krauskopf2022}), and higher federal funds rates (\citealp{Jeff2022})) could have caused ``the perfect storm" in the cryptocurrency market. Moreover, it is relevant to consider that the vicious dependence of the Terra project on the Anchor protocol could have increased its exposure to heterogeneous speculative strategies that occurred simultaneously by chance. We also show how, after 07 May 2022, LUNA is quickly marginalised by the rest of the network's components. This last finding enforces the conjecture that investors considered this collapse a non-structural shock, which is supported by the impossibility of detecting a herding behaviour during the down market.

Finally, to prove our results' robustness, we compare buy/selling dynamics detected on Kraken digital currency exchange with those detected on Binance, highlighting how adverse market actors could have used multiple exchanges to attack/defend the Terra ecosystem.

\section*{Acknowledgements}

The author, A.B. acknowledges Dr. Silvia Bartolucci for her precious support in data retrieving. The author, D.V-T., acknowledges the financial support from the Margarita Salas contract MGS/2021/13 (UP2021-021) financed by the European Union-NextGenerationEU. The author, T.A, acknowledges the financial support from ESRC (ES/K002309/1), EPSRC (EP/P031730/1) and EC (H2020-ICT-2018-2 825215). All the authors acknowledge Cryptocompare for easing the data access. All the authors acknowledge Miranda Zhang for her help in producing graphical representations.

\bibliography{bibliography}

\section*{Supplementary Material}\label{appendix: Supplementary_Material}

\begin{table}[H]
\tiny
\caption*{List of the 61 cryptocurrencies analysed in the current paper. For each asset, the symbol, the name and the corresponding sector according to the taxonomy proposed by \cite{Messari} is reported. There is no consensus on a unique mapping between cryptocurrencies and sectors. The chosen taxonomy is the one adopted by Kraken digital currency exchange.}
\label{tab:cryptocurrencies_list}
\begin{tabular}{@{}ccc@{}}
\toprule
\textbf{Symbol} & \textbf{Name}            & \textbf{Sector}               \\ \midrule
1INCH           & 1inch Network            & Decentralized Exchanges       \\
AAVE            & Aave                     & Lending                       \\
ALGO            & Algorand                 & Smart Contract Platforms      \\
APE             & ApeCoin                  & Other                         \\
ATLAS           & Star Atlas               & Gaming                        \\
ATOM            & Cosmos                   & Smart Contract Platforms      \\
AVAX            & Avalanche                & Smart Contract Platforms      \\
AXS             & Axie Infinity            & Gaming                        \\
BADGER          & Badger DAO               & Decentralized Exchanges       \\
BAL             & Balancer                 & Decentralized Exchanges       \\
BAT             & Basic Attention Token    & Advertising                   \\
BCH             & Bitcoin Cash             & Currencies                    \\
BTC             & Bitcoin                  & Currencies                    \\
CHZ             & Chiliz                   & Payment Platforms             \\
COMP            & Compound                 & Lending                       \\
CQT             & Covalent                 & Data Management               \\
CRV             & Curve DAO Token          & Decentralized Exchanges       \\
DAI             & Dai                      & Stablecoins                   \\
DOGE            & Dogecoin                 & Currencies                    \\
DOT             & Polkadot                 & Other                         \\
DYDX            & dYdX                     & Decentralized Exchanges       \\
ENJ             & Enjin Coin               & Gaming                        \\
ENS             & Ethereum Name Service    & Identity                      \\
ETH             & Ethereum                 & Smart Contract Platforms      \\
FIDA            & Bonfida                  & Decentralized Exchanges       \\
FXS             & Frax Share               & Stablecoins                   \\
GALA            & Gala                     & Gaming                        \\
GMT             & STEPN                    & Other                         \\
GRT             & The Graph                & Data Management               \\
IMX             & Immutable X              & Other                         \\
KIN             & Kin                      & Social Media                  \\
KNC             & Kyber Network Crystal v2 & Decentralized Exchanges       \\
LINK            & Chainlink                & Other                         \\
LRC             & Loopring                 & Decentralized Exchanges       \\
LTC             & Litecoin                 & Currencies                    \\
LUNA            & Terra                    & Smart Contract Platforms      \\
MANA            & Decentraland             & Virtual And Augmented Reality \\
MATIC           & Polygon                  & Scaling                       \\
MKR             & Maker                    & Lending                       \\
OMG             & OMG Network              & Scaling                       \\
PAXG            & PAX Gold                 & Other                         \\
PERP            & Perpetual Protocol       & Derivatives                   \\
REN             & Ren                      & Interoperability              \\
RNDR            & Render Token             & Shared Compute                \\
SAND            & The Sandbox              & Gaming                        \\
SHIB            & Shiba Inu                & Other                         \\
SNX             & Synthetix                & Derivatives                   \\
SOL             & Solana                   & Smart Contract Platforms      \\
SPELL           & Spell Token              & Lending                       \\
SRM             & Serum                    & Decentralized Exchanges       \\
STORJ           & Storj                    & File Storage                  \\
SUSHI           & SushiSwap                & Decentralized Exchanges       \\
TRX             & TRON                     & Smart Contract Platforms      \\
UNI             & Uniswap                  & Decentralized Exchanges       \\
USDT            & Tether                   & Stablecoins                   \\
UST             & TerraUSD                 & Stablecoins                   \\
WAVES           & Waves                    & Smart Contract Platforms      \\
XRP             & XRP                      & Currencies                    \\
YFI             & yearn.finance            & Asset Management              \\
YGG             & Yield Guild Games        & Gaming                        \\
ZRX             & 0x                       & Decentralized Exchanges       \\ \bottomrule
\end{tabular}
\end{table}

\end{document}